\documentclass[prd,twocolumn,superscriptaddress,nofootinbib]{revtex4}
\usepackage{epsfig}
\usepackage{amssymb}
\usepackage{amsmath}
\usepackage{dcolumn}

\def\etal{\mbox{{\it et al.}}}
\newcommand{\livetime}{\mbox{$418\ \mathrm{days}$}}    
\newcommand{\exposure}{\mbox{6.18}}    
\newcommand{\fidexposure}{\mbox{4.54}}  

\newcommand{\Nback}{\mbox{$4.4\pm0.5$}}
\newcommand{\Nbackfull}{\mbox{$4.4\pm0.4(stat.)\pm0.3(sys.)$}}
\newcommand{\NormSYS}{\mbox{$10\,\%$}}

\newcommand{\NullCons}{\mbox{$98\,\%$}}  
\newcommand{\bestDM}{\mbox{$1.3\times10^{-3}\,\mathrm{eV}^2$}}
\newcommand{\bestSST}{\mbox{$0.90$}}
\newcommand{\lowDM}{\mbox{$7\times10^{-5}\,\mathrm{eV}^2$}}
\newcommand{\highDM}{\mbox{$5\times10^{-2}\,\mathrm{eV}^2$}}
\newcommand{\lowSST}{\mbox{$0.2$}}

\newcommand{\NobsGood}{\mbox{$77$}}
\newcommand{\NobsLowRes}{\mbox{$30$}}
\newcommand{\Nobs}{\mbox{$107$}}
\newcommand{\NobsGoodUp}{\mbox{$28$}}
\newcommand{\NobsGoodDown}{\mbox{$49$}}

\newcommand{\Nexpnoosc}{\mbox{$127\pm13$}}       
\newcommand{\NexpnooscGood}{\mbox{$90\pm9$}}   
\newcommand{\NexpnooscLowRes}{\mbox{$37\pm4$}} 
\newcommand{\NexpSK}{\mbox{$96\pm10$}}      
\newcommand{\NexpSKGood}{\mbox{$68\pm7$}}   
\newcommand{\NexpSKLowRes}{\mbox{$28\pm3$}} 

\newcommand{\Bfield}{1.3\,T}

\newcommand{\Nobsnu}{\mbox{$34$}}
\newcommand{\Nobsnubar}{\mbox{$18$}}
\newcommand{\NobsQambig}{\mbox{$25$}}

\newcommand{\Nexpnunoosc}{\mbox{$42\pm4$}}    
\newcommand{\Nexpnubarnoosc}{\mbox{$23\pm2$}}  
\newcommand{\NexpnooscQambig}{\mbox{$26\pm3$}} 

\newcommand{\NexpnuSK}{\mbox{$31\pm3$}}       
\newcommand{\NexpnubarSK}{\mbox{$17\pm2$}}    
\newcommand{\NexpSKQambig}{\mbox{$20\pm2$}}

\newcommand{\depth}{\mbox{$2070$}}

\newcommand{\rupdown}{\mbox{$0.62$}}     
 
\newcommand{\rupdownstatplus}{0.19} 
\newcommand{\rupdownstatminus}{0.14} 
\newcommand{\rupdownsys}{\mbox{$0.02$}}

\newcommand{\rnubarnu}{\mbox{$0.96$}}     
 
\newcommand{\rnubarnustatplus}{0.38} 
\newcommand{\rnubarnustatminus}{0.27} 
\newcommand{\rnubarnusys}{\mbox{$0.15$}}
\newcommand{\rnubarnuexpsys}{\mbox{$0.06$}}
\newcommand{\rnubarnufluxsys}{\mbox{$0.04$}}
\newcommand{\rnubarnuxsecsys}{\mbox{$0.13$}}

\newcommand{\nunubarrat}{\mbox{$0.53$}}

\newcommand{\nunubarratstatplus}{0.21}
\newcommand{\nunubarratstatminus}{0.15}
\newcommand{\nunubarratsys}{\mbox{$0.03$}}

\newcommand{\nue}{\mbox{${\nu_e}$}}
\newcommand{\numu}{\mbox{${\nu_\mu}$}}
\newcommand{\nutau}{\mbox{${\nu_\tau}$}}
\newcommand{\Emu}{\mbox{${\mathrm{E}_\mu}$}}

\newcommand{\numubar}{\mbox{${\overline{\nu}_\mu}$}}

\newcommand{\nuebar}{\mbox{${\overline{\nu}_e}$}}

\newcommand{\tracez}{\mbox{$\Delta_Z$}}
\newcommand{\tposmean}{\mbox{$\langle \Delta_\mathrm{UV}\rangle$}}
\newcommand{\tposrms}{\mbox{$\langle\Delta_\mathrm{UV}^2\rangle^{\frac{1}{2}}$}}
\newcommand{\thetazen}{\mbox{$\theta_\mathrm{zen}$}}
\newcommand{\Qvtx}{\mbox{$Q_\mathrm{{vtx}}$}}
\newcommand{\RMSUP}{\mbox{$\mathrm{RMS}_{\mathrm{UP}}$}}
\newcommand{\RMSDOWN}{\mbox{$\mathrm{RMS}_{\mathrm{DOWN}}$}}

\newcommand{\coszen}{\mbox{$\cos\theta_z$}}

\newcommand{\Deltamsq}{\mbox{$\Delta\mathrm{m}_{23}^2$}}
\newcommand{\Sinsq}{\mbox{$\sin^22\theta_{23}$}}
\newcommand{\Deltamsqgen}{\mbox{$\Delta\mathrm{m}_{gen}^2$}}
\newcommand{\Sinsqgen}{\mbox{$\sin^22\theta_{gen}$}}
\newcommand{\sigmaLoE}{\mbox{$\sigma_{\log{(L/E)}}$}}
\newcommand{\GeV}{\mbox{$\mathrm{GeV}$}}
\newcommand{\eV}{\mbox{$\mathrm{eV}$}}

\begin{document}

\begin{flushleft}
 {\large FERMILAB-PUB-05-525} 
\end{flushleft}
\bigskip\bigskip

\title{\boldmath
 First Observations of Separated Atmospheric $\numu$ and 
 $\numubar$ \\ Events in the MINOS Detector}

\newcommand{\Cambridge}{Dept. of Physics, Cavendish Laboratory, Univ. of Cambridge, Madingley Road, Cambridge CB3 0HE, UK}
\newcommand{\FNAL}{Fermi National Accelerator Laboratory, Batavia, IL 60510}
\newcommand{\RAL}{Rutherford Appleton Laboratory, Chilton, Didcot, Oxfordshire, OX11 0QX, UK}
\newcommand{\Caltech}{Lauritsen Lab, California Institute of Technology, Pasadena, CA 91125}
\newcommand{\ANL}{Argonne National Laboratory, Argonne, IL 60439}
\newcommand{\Athens}{Department of Physics, University of Athens, GR-15771 Athens, Greece}
\newcommand{\NTUAthens}{Dept. of Physics, National Tech. Univ. of Athens, GR-15780 Athens, Greece}
\newcommand{\Benedictine}{Physics Dept., Benedictine University, Lisle, IL 60532}
\newcommand{\BIMC}{Dept. of Rad. Oncology, Beth Israel Med. Center, New York, NY 10003}
\newcommand{\BNL}{Brookhaven National Laboratory, Upton, NY 11973}
\newcommand{\Beijing}{Inst. of High Energy Physics, Chinese Academy of Sciences, Beijing 100039, China}
\newcommand{\CdF}{APC -- Coll\`{e}ge de France, 11 Place Marcelin Berthelot, F-75231 Paris Cedex 05, France}
\newcommand{\Columbia}{Physics Department, Columbia University, New York, NY 10027}
\newcommand{\GEHealth}{GE Healthcare, Florence SC 29501}
\newcommand{\DOE}{Div. of High Energy Physics, U.S. Dept. of Energy, Germantown, MD 20874}
\newcommand{\Harvard}{High Energy Physics Lab, Harvard University, Cambridge, MA 02138}
\newcommand{\HolyCross}{Holy Cross College, Notre Dame, IN 46556}
\newcommand{\IIT}{Physics Division, Illinois Institute of Technology, Chicago, IL 60616}
\newcommand{\Indiana}{Physics Department, Indiana University, Bloomington, IN 47405}
\newcommand{\ITEP}{High Energy Exp. Physics Dept., Inst. of Theor. and Exp. Physics, 
  B. Cheremushkinskaya, 25, 117218 Moscow, Russia}
\newcommand{\JMU}{Physics Dept., James Madison University, Harrisonburg, VA 22807}
\newcommand{\JINR}{Joint Inst. for Nucl. Research, Dubna, Moscow Region, RU-141980, Russia}
\newcommand{\LASL}{Nucl. Nonprolif. Div., Threat Reduc. Dir., Los Alamos National Laboratory, Los Alamos, NM 87545}
\newcommand{\LBL}{Physics Div., Lawrence Berkeley National Laboratory, Berkeley, CA 94720}
\newcommand{\Lebedev}{Nuclear Physics Dept., Lebedev Physical Inst., Leninsky Prospect 53, 117924 Moscow, Russia}
\newcommand{\LLL}{Lawrence Livermore National Laboratory, Livermore, CA 94550}
\newcommand{\MIT}{Lincoln Laboratory, Massachusetts Institute of Technology, Lexington, MA 02420}
\newcommand{\Minnesota}{University of Minnesota, Minneapolis, MN 55455}
\newcommand{\Crookston}{Math, Science and Technology Dept., Univ. of Minnesota -- Crookston, Crookston, MN 56716}
\newcommand{\Duluth}{Dept. of Physics, Univ. of Minnesota -- Duluth, Duluth, MN 55812}
\newcommand{\Oxford}{Sub-dept. of Particle Physics, Univ. of Oxford,  Denys Wilkinson Bldg, Keble Road, Oxford OX1 3RH, UK}
\newcommand{\PSU}{Dept. of Physics, Pennsylvania State Univ., University Park, PA 16802}
\newcommand{\Pittsburgh}{Dept. of Physics and Astronomy, Univ. of Pittsburgh, Pittsburgh, PA 15260}
\newcommand{\IHEP}{Inst. for High Energy Physics, Protvino, Moscow Region RU-140284, Russia}
\newcommand{\RoyalH}{Physics Dept., Royal Holloway, Univ. of London, Egham, Surrey, TW20 0EX, UK}
\newcommand{\Carolina}{Dept. of Physics and Astronomy, Univ. of South Carolina, Columbia, SC 29208}
\newcommand{\SLAC}{Stanford Linear Accelerator Center, Stanford, CA 94309}
\newcommand{\Stanford}{Department of Physics, Stanford University, Stanford, CA 94305}
\newcommand{\Sussex}{Dept. of Physics and Astronomy, University of Sussex, Falmer, Brighton BN1 9QH, UK}
\newcommand{\TexasAM}{Physics Dept., Texas A\&M Univ., College Station, TX 77843}
\newcommand{\Texas}{Dept. of Physics, Univ. of Texas, 1 University Station, Austin, TX 78712}
\newcommand{\TechX}{Tech-X Corp, Boulder, CO 80303}
\newcommand{\Tufts}{Physics Dept., Tufts University, Medford, MA 02155}
\newcommand{\UNICAMP}{Univ. Estadual de Campinas, IF-UNICAMP, CP 6165, 13083-970, Campinas, SP, Brazil}
\newcommand{\USP}{Inst. de F\'{i}sica, Univ. de S\~{a}o Paulo,  CP 66318, 05315-970, S\~{a}o Paulo, SP, Brazil}
\newcommand{\UCL}{Dept. of Physics and Astronomy, University College London, Gower Street, London WC1E 6BT, UK}
\newcommand{\Washington}{Physics Dept., Western Washington Univ., Bellingham, WA 98225}
\newcommand{\WandM}{Dept. of Physics, College of William \& Mary, Williamsburg, VA 23187}
\newcommand{\Wisconsin}{Physics Dept., Univ. of Wisconsin, Madison, WI 53706}
\newcommand{\deceased}{Deceased.}

\affiliation{\ANL}
\affiliation{\Athens}
\affiliation{\Benedictine}
\affiliation{\BNL}
\affiliation{\Caltech}
\affiliation{\Cambridge}
\affiliation{\UNICAMP}
\affiliation{\Beijing}
\affiliation{\CdF}
\affiliation{\Columbia}
\affiliation{\FNAL}
\affiliation{\Harvard}
\affiliation{\IIT}
\affiliation{\Indiana}
\affiliation{\IHEP}
\affiliation{\ITEP}
\affiliation{\JMU}
\affiliation{\JINR}
\affiliation{\Lebedev}
\affiliation{\LLL}
\affiliation{\Minnesota}
\affiliation{\Duluth}
\affiliation{\Oxford}
\affiliation{\Pittsburgh}
\affiliation{\RAL}
\affiliation{\USP}
\affiliation{\Carolina}
\affiliation{\Stanford}
\affiliation{\Sussex}
\affiliation{\TexasAM}
\affiliation{\Texas}
\affiliation{\Tufts}
\affiliation{\UCL}
\affiliation{\Washington}
\affiliation{\WandM}
\affiliation{\Wisconsin}

\author{P.~Adamson}
\affiliation{\FNAL}
\affiliation{\UCL}
\affiliation{\Sussex}

\author{T.~Alexopoulos}
\altaffiliation[Now at\ ]{\NTUAthens .}
\affiliation{\Wisconsin}

\author{W.~W.~M.~Allison}
\affiliation{\Oxford}

\author{G.~J.~Alner}
\affiliation{\RAL}

\author{K.~Anderson}
\affiliation{\FNAL}

\author{C.~Andreopoulos}
\affiliation{\RAL}
\affiliation{\Athens}

\author{M.~Andrews}
\affiliation{\FNAL}

\author{R.~Andrews}
\affiliation{\FNAL}

\author{C.~Arroyo}
\affiliation{\Stanford}

\author{S.~Avvakumov}
\affiliation{\Stanford}

% ...

\collaboration{The MINOS Collaboration}
\noaffiliation

\begin{abstract}
The complete 5.4 kton MINOS far detector has been taking data since
the beginning of August 2003 at a depth of 2070 meters water-equivalent 
in the Soudan mine, 
Minnesota. This paper presents the first MINOS observations of \numu\ and 
\numubar\ charged-current atmospheric neutrino interactions based on an 
exposure of $\livetime$. The 
ratio of upward to downward-going events in the data is compared to the
Monte Carlo expectation in the absence of neutrino oscillations giving: 
\begin{eqnarray*}
 R_{\mathrm{up/down}}^{\mathrm{data}}/R_{\mathrm{up/down}}^{\mathrm{MC}}  &=&  
   \rupdown^{+\rupdownstatplus}_{-\rupdownstatminus} (stat.) \pm \rupdownsys (sys.).
\end{eqnarray*}
An extended maximum likelihood analysis of the 
observed $L/E$ distributions excludes the null hypothesis of no neutrino
oscillations at the $\NullCons$ confidence level. 
Using the curvature of the observed muons in the \Bfield\ MINOS magnetic field 
$\numu$ and $\numubar$ interactions are separated. 
The ratio of $\numubar$ to $\numu$ events in the data is compared to the
Monte Carlo expectation assuming neutrinos and anti-neutrinos oscillate in same
manner giving: 
\begin{eqnarray*}
 R_{\mathrm{\numubar/\numu}}^{\mathrm{data}}/R_{\mathrm{\numubar/\numu}}^{\mathrm{MC}} & = & 
   \rnubarnu^{+\rnubarnustatplus}_{-\rnubarnustatminus} (stat.) \pm \rnubarnusys (sys.),
\end{eqnarray*}
where the errors are the statistical and systematic uncertainties.
Although the statistics are limited, this is the first direct observation of  
atmospheric neutrino interactions separately for $\numu$ and $\numubar$.
\end{abstract}

\pacs{}

\maketitle
 
\section{Introduction}

Over the course of the past ten years the deficit of 
muon neutrinos from cosmic-ray showers in the atmosphere 
has been firmly established by the 
Super-Kamiokande experiment\cite{bib:SuperK_results,bib:SuperK,bib:SuperKI,bib:SuperKII,bib:SuperKIII,bib:SuperKPhase1} and confirmed by the
MACRO\cite{bib:MACRO_results} and Soudan~2\cite{bib:Soudan2_results} 
experiments. The favoured interpretation of the data is 
$\numu\leftrightarrow\nutau$ neutrino oscillations.
Recent results from the Super-Kamiokande experiment\cite{bib:SuperK_new}
provide direct evidence for atmospheric neutrino oscillations and
yield best fit oscillation parameters of $(\Deltamsq,\Sinsq) = 
(0.0024\,\eV^2,1.0)$, where $\Deltamsq=|m_3^2-m_2^2|$. 
Results from the K2K experiment\cite{bib:K2K} 
provide further confirmation of the $\numu\leftrightarrow\nutau$ 
oscillation hypothesis.

The 5.4\,kiloton (kt) mass of the recently
constructed  MINOS  (Main Injector Neutrino Oscillation Search) far detector\cite{bib:Adamson} is much less than 
the \mbox{$\sim25\,$kt} fiducial mass of the Super-Kamiokande detector. However, it 
does possess one unique advantage, namely it is the first large 
deep underground detector to have a magnetic field. This allows  
studies of  neutrino flavour oscillations for neutrinos and 
anti-neutrinos separately. 
A separate measurement of $\numu$ and $\numubar$
oscillations could provide constraints on CPT violating 
models\cite{bib:CPT1,bib:CPT2} which have been invoked to  
accommodate simultaneously 
the solar, atmospheric and LSND\cite{bib:LSND} neutrino 
oscillation data. It should be noted that a number of recent studies
have indicated difficulties with the CPT violating models 
(see for example~\cite{bib:CPT3}). Nevertheless, a direct measurement of
$\numu$ and $\numubar$ oscillations is of interest.   
In addition, MINOS is unique in its ability to provide an accurate 
measurement of the neutrino energy and direction for all contained-vertex $\numu$ 
charged-current (CC) interactions.

This paper presents first results  on atmospheric neutrinos from 
the MINOS experiment. Here, only results from $\numu$/$\numubar$ CC 
events with neutrino interaction vertices contained inside the
detector volume are considered; results from events where the neutrino 
interacts in the surrounding rock will be the subject of a 
separate publication. The data used 
were recorded between August 2003 and February 2005 and correspond to 
a livetime of $\livetime$ giving an exposure of $\exposure$ kiloton-years 
($\fidexposure$ kiloton-years fiducial). 
The data are compared to the expectation in the absence of neutrino oscillations and
the favoured hypothesis of $\numu\leftrightarrow\nutau$ oscillations with 
$\Delta m^2_{23}=0.0024$\,eV$^2$ and $\Sinsq=1.0$. 
The first direct results showing
charge separated $\numu$ and $\numubar$ atmospheric neutrino interactions
are presented. 

\section{The MINOS Detector}

The MINOS far detector is located at a depth of 
$\depth$ meters-water-equivalent (mwe) in the Soudan mine, Northern Minnesota.
The far detector is a steel-scintillator sampling calorimeter
consisting of two super-modules (SM) separated by a gap of 
1.1\,m. The detector consists of octagonal planes
of 2.54\,cm thick steel followed by planes of 1\,cm thick extruded polystyrene
scintillator
and a 2\,cm wide air gap. The first and second SMs are comprised of 
248 and 236 scintillator planes respectively. Each SM is magnetized to 
an average value of \Bfield\ by a 15\,kA current loop which runs through
the coil hole along the detector central axis and returns below the detector. 
Each scintillator plane is made up of 192 strips of width 4.1\,cm 
and of length between $3.4-8.0$\,m depending on position in the plane. The strips
in alternating planes are oriented at $\pm45^\circ$ to the 
vertical thereby providing two orthogonal coordinates\footnote{
The MINOS right-handed coordinate system has the $z$-axis defined 
along the detector axis pointing away from Fermilab and the 
$y$-axis vertical. The alternating
scintillator planes provide measurements of the $U$ and $V$
coordinates which are related to $x$ and $y$ by 
$U=\frac{1}{\sqrt{2}}(x+y)$ and $V=\frac{1}{\sqrt{2}}(y-x)$.}.    
The scintillation light is collected using wavelength 
shifting (WLS) fibers embedded within the scintillator strips.
The WLS fibers are coupled to clear 
optical fibers at both ends of a strip and are read out using 16-pixel
multi-anode photomultiplier tubes (PMTs). 
The signals from eight strips, separated by approximately 1\,m 
within the same plane, are optically summed (multiplexed) and read 
out by a single PMT pixel. 
The multiplexing pattern is different for the two sides of the detector,
which, for a single hit, enables the resulting eightfold ambiguity to be 
resolved. For all types of event, the ambiguities are efficiently resolved in
software using additional information from timing and event
topology.

The detector is optimized for detecting beam neutrinos coming from the 
direction of Fermilab. For the study of atmospheric neutrinos the planar
structure presents a particular problem: cosmic-ray muons traveling
almost parallel to the scintillator 
planes can penetrate deep into the detector by
traveling in the steel or air between the planes. To reject this source
of background a scintillator veto shield surrounds the upper part of the 
main detector. The veto shield is constructed from the same scintillator modules
as used in the main detector but with the orientation of strips
aligned along the z-axis.
The veto shield comprises a ``ceiling'' section above the detector, consisting 
of two scintillator layers, and ``wall'' sections along each of the 
two sides of the detector formed from a single scintillator layer.  

\subsection{Data Acquisition and Trigger}

The output signals from
each PMT pixel are digitized and time-stamped (with a 1.5625\,ns precision)
by the VME-based front-end electronics. The signals from the pixels are 
digitized by 14-bit analogue-to-digital  converters (ADC) when the dynode 
signal from the PMT exceeds a programmable threshold, corresponding to
approximately one third of a photo-electron.
 To reduce the data flow, the pedestal corrected signals are only written
 to the DAQ output buffers if two out of thirty-six channels on the same
 readout board are above threshold. These thirty-six channels
 correspond to the readout on one side of the detector from a
 contiguous group of either 20 or 24 planes. The raw data rate
 is approximately 8\,MB\,s$^{-1}$. The raw
 data are transferred to a PC based trigger farm where
 the data are divided into blocks bounded by regions of 100 clock ticks
 (156\,ns) or more where no detector activity has been recorded.
 The primary trigger algorithm, applied to these blocks of data,
 requires there to be activity in at least $4$ planes out of any contiguous
 group of $5$ planes.  The veto shield is read out in the same manner as
 the main detector except that the two out of thirty-six requirement is
 not applied and the dynode threshold is set to a level corresponding to
 approximately one and a half photo-electrons.

The MINOS far detector front-end electronics and data acquisition 
system are described in detail in~\cite{bib:FE} and~\cite{bib:DAQ}.

\subsection{Detector Calibration}

A minimum ionizing particle crossing at normal incidence to a plane 
gives a combined signal of approximately $10$ photo-electrons (PEs) 
registered by the PMTs at the two ends of the strip. The detector is 
calibrated using both a dedicated LED system\cite{bib:LI} and cosmic-ray muons.
The ADC to PE calibration is performed using the LED system and the
cosmic-ray muon sample is then used to give a uniform response across
the detector. From studies of cosmic-ray muons in the MINOS
detector\cite{bib:Hartnell}, the current uncertainty in the PE calibration
is 5\,\%. Cosmic-ray muons are also used to calibrate the recorded times. 
After calibration, a single hit timing resolution of approximately 2.3\,ns 
is achieved. The timing calibration tracks all hardware changes.

\section{Data and Monte Carlo}

The data described in this paper were recorded in the 18 month period
from August 2003 to February 2005. Only data taken when the MINOS far 
detector, including the veto shield, was fully operational are used.
The final data sample corresponds to a livetime of $\livetime$ giving an
exposure of $\exposure$ kiloton-years ($\fidexposure$ kiloton-years fiducial).

The selection of contained-vertex neutrino interactions was optimized using
a GEANT\,3\cite{bib:Geant3} simulation of the MINOS detector. 
For the simulation of atmospheric neutrino events the 3D flux  
calculation of Barr\ \etal\cite{bib:Barr} was used (Bartol 3D).
The NEUGEN3 program\cite{bib:NEUGEN} was used to simulate the neutrino 
interactions (cross sections and hadronic final states). 
The earlier 1D flux calculation from the Bartol group\cite{bib:Bartol} (Bartol 1D) 
and the 3D calculation of Battistoni\ \etal\cite{bib:Battistoni} were 
used to assign systematic
uncertainties.
The response of the MINOS detector to electrons, muons and hadrons has been 
studied in a test beam at the CERN PS using the 12.5 ton MINOS calibration 
detector\cite{bib:caldet}. The test beam detector was constructed and
read out in the same manner as the MINOS far detector. 
The interactions of hadronic particles are modeled with the 
GCALOR package\cite{bib:GCALOR}, which is found to give a reasonable
description of low energy hadronic interactions in the MINOS 
calibration detector\cite{bib:Kordosky}, rather than the default version of
GHEISHA (see \cite{bib:Geant3} and references therein). 
The ``SLAC version''\cite{bib:SLACGHEISHA} of 
GHEISHA, which also provides a reasonable 
description of the test beam data, is used as an alternative 
model for hadronic interactions.
A Monte Carlo (MC) sample of atmospheric neutrino interactions
corresponding to over 1000 live-years was generated and used to 
optimize both the reconstruction algorithms and the event selection 
criteria.
Two large cosmic-ray muon background samples were generated: 
a sample of 19 million events full spectrum 
(corresponding to approximately 280 days livetime) and a further 
2 million events with $\Emu<2\,\GeV$ (corresponding to a livetime of
4.1 years) as lower energy events are an important component
of the cosmic-ray muon background to the contained-vertex atmospheric
neutrino selection.
A 10\,\% uncertainty on the normalisation of the cosmic-ray 
background is assigned. The error reflects the different normalisation obtained 
when normalising to the entire cosmic-ray sample or normalising to just those 
cosmic-ray muons which stop in the detector (these form the main background to
the event selection described below).
It should be noted that for the results 
presented in this paper the cosmic-ray background in the selected event 
sample is estimated from data. The 10\,\% uncertainty in the cosmic-ray normalisation 
is only used when comparing data and Monte Carlo at various stages in the event
selection.

\subsection{Flux Normalization and Systematic Uncertainties}

\label{sec:fluxnorm}

The theoretical prediction for the atmospheric neutrino 
event rate has large uncertainties from the primary cosmic-ray flux,
hadron production models and neutrino interaction cross sections.
The analysis of the Soudan~2 $\nue/\nuebar$ data\cite{bib:Soudan2_results}
indicates that the combined prediction of the Bartol 3D model\cite{bib:Barr} 
and the NEUGEN3\cite{bib:NEUGEN} neutrino cross section model
should be scaled by $0.88\pm0.07$\cite{bib:Soudan2_new}, where the error 
is statistical and it is implicitly assumed that atmospheric electron 
neutrinos are not oscillating. This normalization result is compatible 
with the results from a fit to the Soudan~2 data including oscillations. 
The MINOS and Soudan~2 detectors are located in the same mine 
({\em i.e.} at the same geomagnetic latitude) and both are 
constructed from steel. 
Consequently, for the analysis presented here the 
Soudan~2 scale factor $0.88\pm0.07$ is used to correct the combined 
event rate predictions from the Bartol 3D flux model 
and the NEUGEN3 neutrino interaction model.
An additional 5\,\% uncertainty is added in quadrature to that estimated
by the Soudan~2 collaboration to account for differences
arising from the different energy thresholds ($300$\,MeV in 
the case of Soudan~2 compared to $\sim500$\,MeV for MINOS).
Finally an additional 2.5\,\% uncertainty is 
assigned to account for the different phases in the solar cycle for the 
Soudan~2 and MINOS data sets\cite{bib:BarrPC}; because of the relatively high neutrino 
energy threshold the selected atmospheric neutrino rate in MINOS depends 
only weakly on the phase in the solar cycle. 
The resulting total systematic uncertainty on the
expected event rate is estimated to be 10\,\%.

\section{Event Reconstruction}

\label{sect:reco}

The MINOS detector is optimized for beam neutrinos originating from
Fermilab. Due to the curvature of the Earth, beam neutrinos enter the detector 
from below the horizontal at an angle of $3.3^\circ$ with respect to the $z$-axis. 
The standard MINOS reconstruction software has been developed for these events. 
The analysis presented here uses reconstruction software optimized
for atmospheric neutrinos\cite{bib:Blake}. 

\begin{figure}
 \includegraphics[width=\columnwidth]{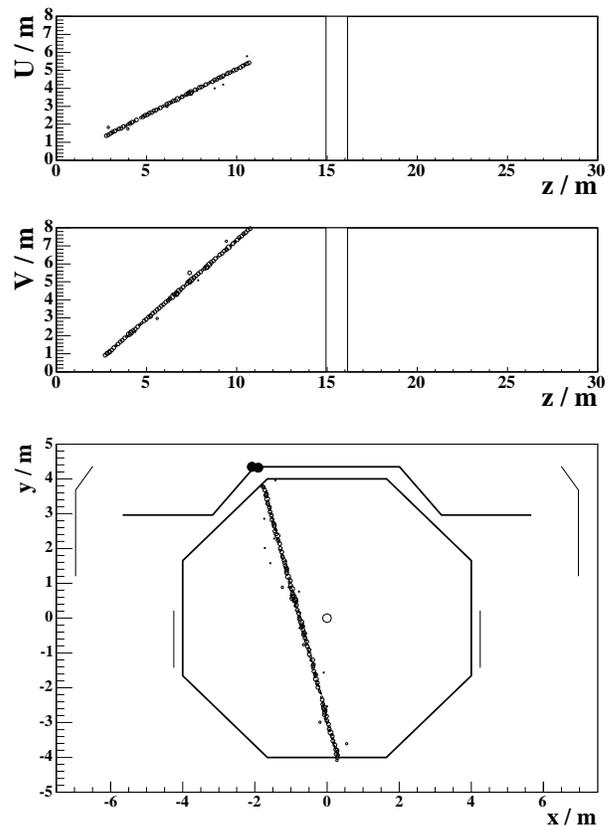}
 \caption{\label{fig:MuonEvent} An example of a cosmic-ray muon event 
          in the MINOS
          far detector. The detector readout corresponds to the 
          two orthogonal $U-z$ and $V-z$ views. The size of the points
          gives an indication of the pulse height for 
          each scintillator strip hit. The large dark points shown in the
          $x-y$ view indicate in-time activity in the veto shield. }  
\end{figure}

The first stage of the event reconstruction
removes the eightfold ambiguity in the association 
of raw hits to strips. This is performed utilizing
information from both strip ends. For cosmic-ray muons, an average of
$99\,\%$ of the recorded pulse height is associated with the correct strip. 
At this stage the data are in the form of two 2D event views
$U-z$ and $V-z$. An example event display of a cosmic-ray muon is
shown in Figure~\ref{fig:MuonEvent}. 
Tracks and showers are
reconstructed independently in each view; the two views are
then matched to obtain a three-dimensional event.
For cosmic-ray events that leave hits in both the veto shield and 
main MINOS far detector, the root-mean-square (rms) difference
in times recorded in veto shield and the detector is $4$\,ns, allowing
association of veto shield hits (indicated in 
Figure~\ref{fig:MuonEvent}) to activity in the main detector. 

A charged-current muon neutrino event is in general 
reconstructed as a muon track 
and a hadronic shower. A typical 1\,GeV muon will traverse approximately
25 planes at normal incidence. Reconstructed tracks are required 
to consist of at least 8 planes (corresponding to a minimum energy of 
0.4\,GeV). For muons which start and stop within the detector volume
the muon momentum is determined from range with a resolution of
approximately 
$(\sigma_p/{p})^2 = 0.06^2 + 
(0.045/{p})^2$  
for muons traveling at normal incidence to the detector planes
(where $p$ is measured in GeV$/c$).
The first term is dominated by fluctuations in energy loss and the second
is dominated by sampling. 
For events where the muon exits the detector,
the muon momentum is obtained from the
curvature of the track in the magnetic field. For the selected CC 
atmospheric $\numu$/$\numubar$ interactions, where the momentum is
determined from curvature, the average 
momentum resolution is approximately
$\sigma^2_{1/p} = [0.1^2 +  
(0.3/p)^2]\,\mathrm{GeV}^{-2}$ (where $p$ is measured in GeV$/c$).
However, the resolution obtained from individual events depends strongly 
on how much of the trajectory of the muon is observed before it exits the
detector and on the orientation of the trajectory relative to the local
magnetic field. 
The hadronic energy is obtained by summing the pulse height in a 
shower which is spatially associated with the start of the track.
The energy scale is obtained from Monte Carlo using the 
GCALOR\cite{bib:GCALOR} model of hadronic showers, which from
the test beam results is found to provide a good description 
of the detector 
response to single $\pi^\pm$ and protons\cite{bib:Kordosky}. 
The hadronic energy resolution is approximately 
$\sigma_{{E}}/{E} \sim 0.55/\sqrt{{E}}$, where $E$ is measured in
GeV.

\begin{figure}
 \includegraphics[width=\columnwidth]{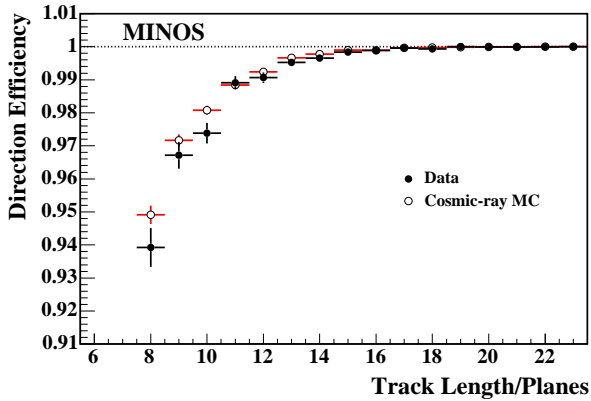}
 \caption{\label{fig:DirectionEff} 
          The efficiency for correctly reconstructing stopping
          muon events as downward-going as a function of number of
          planes in the reconstructed track.}
\end{figure}

For the study of atmospheric neutrinos it is necessary to 
determine whether the reconstructed track is upward or downward going. 
A relativistic normal incidence particle traverses ten planes
in approximately 2\,ns which, when compared to the single hit resolution
of 2.3\,ns, is sufficient to identify the direction of
most selected events with little ambiguity.
The sense of the direction of muon tracks is determined by comparing the hit 
times along the reconstructed track with the hypotheses that it is either upward or 
downward going (assuming that the particle is traveling at the speed of light). 
The rms deviations 
of hit times about each of the two hypotheses are calculated, 
\RMSUP\ and \RMSDOWN. 
The hypothesis with the smallest rms is chosen.
In addition, the magnitude of $\RMSUP-\RMSDOWN$ provides a 
measure of the quality of the direction determination.
To test the performance of the algorithm a sample of
stopping cosmic-ray muons is used (all of which are traveling
downward). Figure~\ref{fig:DirectionEff} shows a comparison of the 
data and Monte Carlo efficiencies for correctly identifying a stopping muon 
as downward-going as a function of the number of planes the track crosses. 
The average efficiency is above 94\,\% for even the shortest tracks and rapidly
increases to better than 99\,\% for events with hits in 12 planes. 
The efficiency in data agrees with that from 
Monte Carlo to better than 1\,\%. 

\begin{figure}
 \includegraphics[width=\columnwidth]{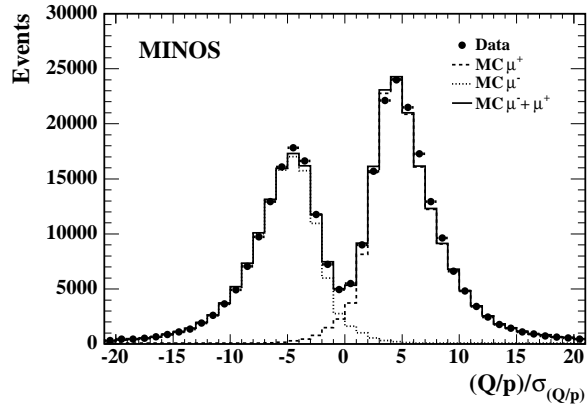}
 \caption{\label{fig:StoppersCharge} The reconstructed distribution of $(Q/p) / \sigma_{(Q/p)}$ for 
     stopping muon events in data and Monte Carlo.}
\end{figure}

The curvature of $\mu^+$/$\mu^-$ tracks 
in the magnetic field allows the charge sign to be determined.
Figure~\ref{fig:StoppersCharge} shows the distribution of  the 
reconstructed charge divided by momentum, $Q/p$, divided by its error, 
for cosmic-ray muons that stop in the detector. Two peaks, corresponding to $\mu^-$ and
$\mu^+$ events, are clearly seen. 
The
widths of the two peaks in data and MC agree to better than 2.5\,\%.
For the event samples considered here, the $\mu^+$/$\mu^-$ charge is 
cleanly identified over the approximate momentum range $1-10$\,\GeV.
The efficiency decreases for low momentum tracks due to the limited number
of planes crossed. For high momentum tracks, which typically
leave the detector, the charge identification efficiency decreases as
only the limited curvature at the start of the track is measured. 

\section{Event Selection}

At a depth of $\depth$ mwe 
the cosmic-ray muon rate is approximately 50000 events per day 
in the MINOS detector. This rate should be
compared to the expected signal rate of $0.54\pm0.05$ atmospheric 
CC
$\numu/\numubar$ interactions per day\footnote{The signal rate of
$0.54\pm0.05$ (no oscillations) corresponds to $\numu/\numubar$ CC interactions where the
muon deposits energy in at least eight planes (before fiducial cuts).}, where the uncertainty is from the
$\NormSYS$ uncertainty in the expected event rate (discussed
in Section~\ref{sec:fluxnorm}). 
In order to achieve a signal-to-background ratio of 
ten-to-one  it is necessary to identify the signal events efficiently  
whilst reducing the background by a factor of $10^6$. 
The event selection is designed to identify both fully-contained (FC) and
partially-contained (PC) $\numu/\numubar$ events. In FC events
the entire event is contained within the fiducial volume. In 
PC events the neutrino vertex is within the fiducial volume but the
produced muon exits the detector.

\subsection{Preselection}

Candidate CC $\numu$ neutrino interactions are required to have a 
reconstructed track passing some basic quality requirements. 
The majority of the background is rejected by event containment 
requirements which are applied at both the hit and reconstructed track level.
The sense of the track direction (up/down) 
is determined from timing as described previously. 
The start of the track, which is considered to be the neutrino interaction 
vertex, is required to lie within the detector fiducial volume. The 
fiducial volume is defined as the octagonal region which is at least 50\,cm 
from the detector edges in the $xy$ plane and at least 5 planes from
the start and end of either SM. In addition, the region within 
40\,cm of the axis of the coil hole, which has a diameter of 25\,cm, is excluded from the fiducial 
volume. This cut is enlarged
to 1\,m in the first and last 10 planes of the detector.
The event sample is sub-divided into FC and PC
events depending on whether the end of the track also lies within the fiducial 
region. 

Event containment cuts are also made at the hit level to reduce the sensitivity
to possible reconstruction errors where not all hits are correctly
associated to the reconstructed track. For this purpose, the
fiducial volume requirement of 50\,cm from the detector edges is relaxed
to 30\,cm. In order to apply the containment cuts at the hit level it is
necessary to convert the two-dimensional coordinates of a single hit
into a point in space. This conversion is achieved by using the mean value of
the orthogonal (U/V) coordinate in the surrounding two planes.  
Hits outside the fiducial volume are then assigned to the nearest octagonal 
edge/SM end. Edges/ends with summed pulse height equivalent to more than six PEs 
outside the fiducial region are tagged as being uncontained. Candidate
FC (PC) events are required to have no (one) such region.
The containment cuts reject approximately 99.9\,\% of the cosmic-ray 
background whilst retaining 77\,\% of CC $\numu/\numubar$ interactions
in the detector volume. The inefficiency for signal events is primarily
a fiducial effect; the containment cuts retain 99\,\% of
CC $\numu/\numubar$ interactions in the fiducial region which 
produce a muon which spans at least six planes.

Candidate FC events are required to have a reconstructed track 
consisting of hits in at least eight planes. 
The PC event selection criteria are optimized separately  
for upward and downward-going events as the backgrounds for the 
two categories are very
different.
To ensure the track direction is well determined, candidate
PC events are required to have a track of at least 1\,m in length and 
which consists of hits in at least ten planes. 

\subsection{Fully Contained and Downward Partially Contained Event Selection}

The dominant backgrounds in the FC and downward PC
samples arise from steep cosmic-ray muons 
which enter the detector at small angles to the detector planes. 
By traveling in the steel or air between the scintillator planes, 
such events can penetrate a significant distance into 
the fiducial volume before leaving a detectable signal. 
The selection of FC and downward-traveling PC 
CC $\numu$/$\numubar$ interactions aims to greatly reduce
this background and proceeds in four stages:
\begin{enumerate}
  \item[i)] {\bf Cosmic-Ray Rejection (Trace cut):}
     The reconstructed track is 
     extrapolated back to the outside of the detector and the distance
     traversed in the direction perpendicular to the detector planes
     is calculated, $\tracez$. Events with small values of $\tracez$ correspond
     to steep tracks which when extrapolated to the detector edge 
     traverse only a few scintillator planes. Figure~\ref{fig:Trace} shows
     the $\tracez$ distribution for MC cosmic-ray muons and CC $\numu/\numubar$
     interactions. Events are rejected if $\tracez<0.5$\,m. 
     Figure~\ref{fig:Trace} also shows the $\tracez$ distribution for 
     data which is in reasonable agreement with the MC expectation.

\begin{figure}
 \includegraphics[width=\columnwidth]{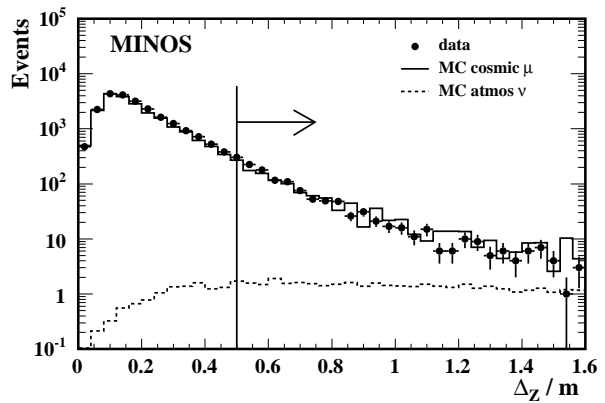}
 \caption{\label{fig:Trace} 
          The reconstructed $\tracez$ distribution for events passing the 
          containment cuts. The solid histogram indicates the MC expectation
          for cosmic-ray background. The points with error bars show the
          observed data. The dashed histogram indicates the expected
          distribution for atmospheric neutrino events (without oscillations).}
\end{figure}

  \item[ii)] {\bf Event Topology:}
     About half of the remaining background consists of cosmic-ray muon 
     tracks that bend in the magnetic field and turn over in the $z$ direction.
     Such events will leave hits in two separate positions in a particular plane.
     In addition, these events typically have large pulse height  
     in the plane where the muon turns around in $z$. This category of background event 
     is rejected using the pulse-height weighted deviations of 
     the hits in the $U-z$ and $V-z$ views from the fitted track. 
     The pulse-height weighted mean, $\tposmean$, and the pulse-height weighted rms 
     deviation of hits from the track, $\tposrms$, are calculated. 
     Events are rejected if there is large scatter about the track,
     $\tposrms>0.5$\,m, or if the pulse-height weighted mean deviation from the
     track lies significantly
     above the reconstructed track, $\tposmean>0.25\,$m. These empirically
     determined cuts are
     applied separately to the hits in both the $U-z$ and $V-z$ views.
     In addition, the event vertex is defined as the first hit on the track 
     taking the highest end (largest $y$) as the start of the track.
     The maximum displacement from the event vertex of the hit strips which lie
     within $\pm4$ planes of the event vertex is found, $\Delta_R^{max}$. 
     Events are rejected if $\Delta_R^{max}>1.25$\,m.

  \item[iii)] {\bf Vertex Pulse Height/Direction:}
     After the topology cut, the signal-to-background ratio is 
     approximately 1:5. The remaining background consists of steep cosmic-ray 
     muons
     which travel nearly parallel to the scintillator planes
     and therefore tend to give a large pulse height signal in a 
     single plane near the beginning of the track.
     These events are often poorly reconstructed due to the difficulties
     of reconstructing tracks for events at small angles to the
     detector planes. Figure~\ref{fig:VertexQvsZenith} shows, for
     signal and background, the total pulse height in the event vertex region, 
     $\Qvtx$, 
     plotted against the cosine of
     reconstructed zenith angle at the highest end of the track\footnote{The zenith 
     angle, $\thetazen$,
     is defined as $\pi$ minus the angle between the reconstructed track 
     direction and the local
     vertical (the $y$-axis). Negative values of the cosine of the
     zenith angle correspond to tracks 
     which are reconstructed as upward-going.} and the modulus of reconstructed
     track direction cosine with respect to the $z$ axis, $|\cos\theta_z|$.
     The vertex pulse height is defined as the maximum number of PEs observed
     in a single plane within $\pm4$ planes of the event vertex (defined above).
     The background is characterized by being steep and having large $\Qvtx$.
     Events are rejected if they have $\Qvtx>300$\,PEs. Steep events, 
     defined as having $|\cos\thetazen|>0.7$ and
     $|\cos\theta_z|<0.5$, are required to satisfy $\Qvtx<100$\,PEs.
     The above event charge/direction 
     cuts are not applied to events with track lengths of greater than 20
     detector planes, as the steep background events tend to cross relatively
     few planes.  
      
\begin{figure}[htb]
  \includegraphics[width=\columnwidth]{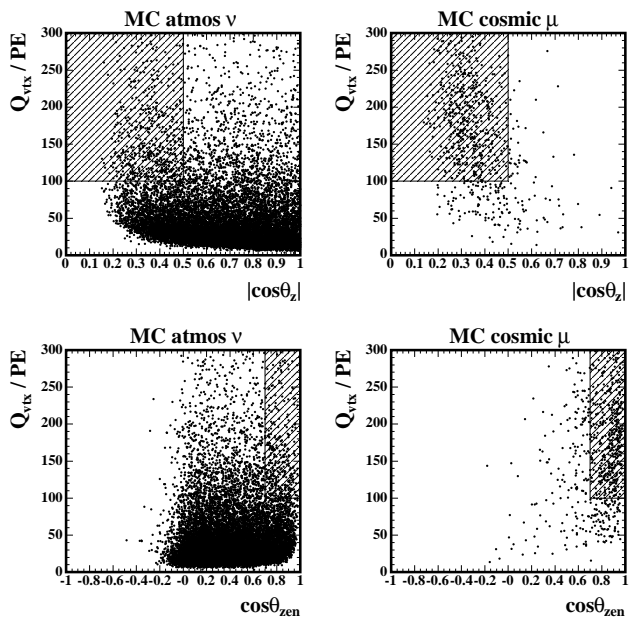}
 \caption{\label{fig:VertexQvsZenith} 
   The MC distributions of the vertex pulse height, $\Qvtx$, plotted against
   the modulus of the cosine of the angle with respect to the detector
   z-axis, $|\cos\theta_z|$ and
   the cosine of the reconstructed zenith angle, $\cos\thetazen$. 
   Plots are shown for both the cosmic-ray muon
   background and for the atmospheric neutrino signal for all events
   passing the containment cuts. The hatched areas represent the 
   regions rejected by the ``Vertex Pulse Height/Direction'' cuts.}
\end{figure}

\item[iv)] {\bf Veto Shield:} 
     The cuts listed above result in a signal-to-background 
     ratio of approximately $1:2$. Additional background
     is removed by rejecting events with activity in the veto shield 
     within a $\pm100$\,ns window around the event time, resulting in
     a signal-to-background ratio of approximately $20:1$.
\end{enumerate}

\subsection{Upward Partially Contained Events}

\begin{figure}
\includegraphics[width=\columnwidth]{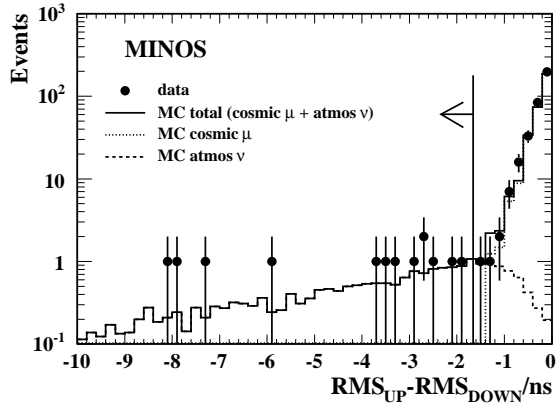}
 \caption{The distribution of $\RMSUP-\RMSDOWN$ for events passing
    all other cuts in the upward partially contained event selection.
    The data are shown by the points with error bars; the total Monte
    Carlo expectation is shown by the solid histogram with the expected 
    atmospheric neutrino 
    contribution (no oscillation) shown by the dashed histogram.  
    The cut is indicated by the arrow.
 \label{fig:timingrms}}
\end{figure}

The background to the upward-going PC event
selection is dominated by cosmic-ray muons which stop in the
detector and are reconstructed as upward rather than downward-going. 
The cuts to remove this source of background are based
on timing information and identify events which are
unambiguously upward-going. The event selection cuts fall into two
categories: 
\begin{enumerate}
  \item[i)] {\bf Event Topology:}
    For the upward PC selection, the relatively small 
    number of badly reconstructed events passing the preselection are removed
    using a subset of the topology cuts employed in the FC/downward
    PC analysis. Events are rejected if $\Qvtx>300$\,PEs or
    $\Delta_R^{max}>1.25$\,m.
  \item[ii)] {\bf Track timing rms:}
    The expected times of hits on the track are calculated
    for the hypotheses of an upward-going and a downward-going track. 
    The rms scatter
    of the difference between observed and expected hit times for these two hypotheses
    is used to identify upward-going events. The event is required
    to be consistent with the upward hypothesis and
    significantly more compatible with the upward hypothesis than
    the downward hypothesis: $\RMSUP<4.33$\,ns 
    and $\RMSUP-\RMSDOWN<-1.66$\,ns (these numbers should
    be compared to the single hit timing resolution of 2.3\,ns). 
    Figure~\ref{fig:timingrms} shows the effect of the main timing cut,
    namely $\RMSUP-\RMSDOWN$, for data compared to the Monte Carlo
    expectation. 
\end{enumerate}

\subsection{Performance}

\begin{figure}
\includegraphics[width=\columnwidth]{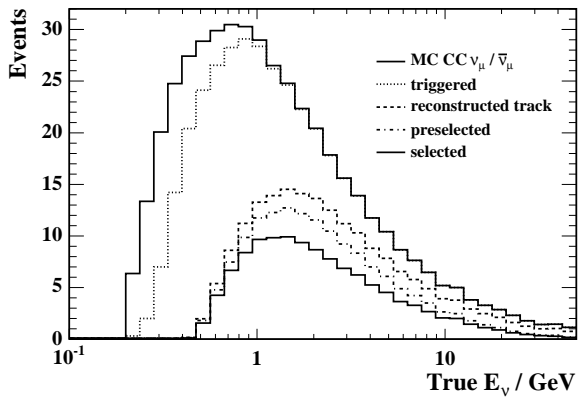}
 \caption{The expected number of CC $\numu$/$\numubar$ events that 
          interact within the detector fiducial volume at the different
          stages in the selection (418 days exposure). In addition to the
          expected energy distribution before selection and the final 
          selected distribution, the expected numbers
          of events are shown at several stages in the analysis: events which 
          pass the trigger requirements; events 
          for which there is a reconstructed track; and events which pass the
          preselection cuts.
 \label{fig:cutflow}}
\end{figure}

The event selection reduces the background from cosmic-ray muons
by a factor of $4\times10^6$. In Monte Carlo the efficiency for CC $\numu$/$\numubar$
neutrino interactions where the interaction occurs within the
fiducial region and the muon traverses eight or more scintillator planes
is 70\,\%.
Figure~\ref{fig:cutflow} shows the expected energy distribution for 
CC $\numu$/$\numubar$ for the various stages in the event 
selection.   
The effective lower limit on the neutrino energy of the selected events is approximately
0.5\,\GeV. For low energy CC neutrino interactions the efficiency is low 
because tracks are only reconstructed if they span at least eight 
detector planes. 

The numbers of events surviving at different stages in event selections
are listed in Table~\ref{tab:fcpcdncutflow}. Reasonable agreement between data and Monte
Carlo is seen at each stage. 
The final veto shield requirement rejects 149 events in data, 
consistent within one standard deviation with the MC expectation of 
$170\pm24$, where the uncertainty is from normalization and 
MC statistics. 
For the results in this paper, the cosmic-ray muon background in the combined
FC and downward PC event sample is estimated from data rather than relying on MC. From 
the 149 events rejected by the veto shield cuts the remaining
background is estimated to be \Nbackfull\ events using the
veto shield efficiency of $97.1\pm0.2\,\%$ (described below) and taking account of
the expected number of neutrino events rejected by the veto shield requirements. 

\begin{table*}[htb]
\centering
\begin{tabular}{|l|c|c|c|c|}
\hline 
Cuts  & Data & \multicolumn{3}{c|}{Monte Carlo Expectation} \\       &      &  total & cosmic muon & $\nu_{\mu}/\bar{\nu}_{\mu}$ CC \\ \hline 
\multicolumn{5}{|c|}{Fully Contained and Downwards Partially Contained} \\ \hline 
Preselection &  $41571$ & $38253\pm3987$ & $38121\pm3987$ &  $125\pm13$ \\ 
Trace        &   $1525$ &  $1513\pm153$  &  $1395\pm153$  &  $112\pm12$ \\ 
Topology     &    $560$ &   $494\pm48$   &   $384\pm47$   &  $104\pm11$ \\ 
Vertex/Direction& $243$ &   $277\pm26$   &   $170\pm24$   &  $102\pm11$ \\ 
 Veto Shield &     $94$ &   $110\pm11$   &   $4.9\pm0.7$  &  $100\pm10$ \\ \hline 
\multicolumn{5}{|c|}{Upward Partially Contained} \\ \hline 
Preselection &    $427$ &   $408\pm47$   &  $384\pm47$    &  $24\pm2$ \\ 
Topology     &    $364$ &   $359\pm42$   &  $336\pm42$    &  $22\pm2$ \\ 
Timing       &     $13$ &    $18\pm2$    &  $<0.36$ (68\% C.L.)&  $17\pm2$ \\   \hline
 \end{tabular}
\caption{The numbers of data events after the different stages of 
the event selection compared with the MC expectation from cosmic-ray 
background events and CC atmospheric $\numu$ and $\numubar$ events. 
The atmospheric neutrino numbers are the MC expectations for no oscillations. 
The uncertainties include MC statistics and
systematic uncertainty on the normalization ($\pm10\,\%$ for the 
cosmic-ray background and $\NormSYS$ for the atmospheric neutrino sample)
and a $3.3\,\%$ systematic uncertainty on the selection efficiency for
CC atmospheric $\numu$ and $\numubar$ events.
The numbers in the total column include neutral current interactions,
$\nue/\nuebar$ CC interactions and interactions of neutrinos in the
surrounding rock. 
\label{tab:fcpcdncutflow}}
\end{table*}

\subsection{Event Selection Systematic Uncertainties}

The systematic uncertainties on the event selection efficiency
and cosmic-ray muon background have been studied in detail. In each 
case the impact of the systematic effect on the MC expectation for the
number of selected events is estimated. In addition, because the selection
is not up-down symmetric, systematic uncertainties are calculated for
the MC expectation for the ratio of upward-going to downward-going 
events. The total uncertainty on the selection efficiency for atmospheric
neutrino events is estimated to be 3.3\,\%. The contribution to the
systematic uncertainty on the up-down ratio from experimental effects is estimated
to be 3.1\,\%.  The contributions to these systematic errors 
are discussed in detail below.

\bigskip
\noindent {\bf{{Veto Shield}}:}
     The efficiency of the veto shield cut is determined directly
     from data in two independent ways. Firstly, a sample of cosmic-ray 
     muons that stop in the detector and have $|\cos\thetazen|>0.5$
     is selected. These events occupy a similar region of
     phase space to the background. The veto shield cut rejects 
     $97.06\pm0.03$\,\% of this sample.
     A second estimate of
     the veto shield efficiency is obtained by relaxing the event selection
     cuts until the sample is dominated by background ({\em i.e.} an expected
     signal fraction of less than 2\,\%). The veto shield cut rejects
     $96.2\pm0.2$\,\% of this sample which, when the expected signal (assuming 
     $\Delta m^2_{23}=0.0024\,\mathrm{eV}^2$) is taken into account, 
     leads to an estimated veto shield efficiency of $97.3\pm0.2$\,\%.
     From these two tests the veto shield efficiency is estimated to
     be $97.1\pm0.2$\,\%, where the central value is taken from
     the high statistics stopping muon sample and a systematic error
     of $0.2$\,\% is added reflecting the difference between the two 
     methods.

     The fraction of signal events rejected due to accidental coincidences
     with hits in the veto shield is estimated by overlaying veto shield
     hits obtained from special minimum bias data taking runs 
     onto Monte Carlo atmospheric
     neutrino events. The estimated fraction of signal events 
     rejected due to spurious veto shield hits is $2.2\pm0.4\,\%$, 
     where the error represents systematic time dependent variations.
     In addition, from Monte Carlo studies it is estimated that 
     $0.3\pm0.1$\,\% of the selected signal downward-going events will be 
     rejected 
     due to hits in the veto shield associated with 
     the neutrino interaction.

\bigskip
\noindent {\bf{{Hadronic Response}}:}
The event selection efficiency depends on the detector response to
hadrons and consequently the hadronic interaction model used. Comparisons
of GCALOR and GHEISHA show no evidence for any significant 
difference in overall selection efficiency or reconstructed up-down
ratio. The GCALOR model is found to provide a good description of the 
response of the detector to single hadrons.
Systematic errors of 2.5\,\% on the selection efficiency and
3.0\,\% on the up-down ratio are assigned; in both cases the estimates 
reflect the Monte Carlo statistical precision of the comparison.

\bigskip
\noindent {\bf{{Scintillator Light Calibration}}:}
 The overall calibration of the MINOS far detector is currently known
  to 5\,\%. The MC response is tuned to agree with cosmic-ray muon data and
 has a corresponding 5\,\% uncertainty. Because the selection cuts
 use pulse height information, this leads to systematic errors of
 0.6\,\% on the selection efficiency and 0.3\,\% on the up-down ratio.

\bigskip
\noindent {\bf{{Timing Calibration/Resolution}}:}
The timing calibration for each scintillator strip is determined from
data in a manner that tracks hardware changes. The uncertainty on the
timing calibration for the individual strips is $0.3$\,ns, {\em i.e.} 
significantly less than the single
hit resolution of 2.3\,ns. The effect on the selection is negligible.
A more significant effect is that the single hit resolution in MC is
better than that in the data, 2.2\,ns compared to 2.3\,ns.
This is due to an incomplete simulation of the electronics readout.
For this reason the times of the Monte 
Carlo hits are smeared by a Gaussian of width 0.7\,ns. The difference between the
the selection efficiencies before and after this smearing are compared. 
The overall selection efficiency for the smeared MC is reduced by 1.0\,\%.
The effect on the up-down ratio is small (0.1\,\%). These differences
are used as estimates of the systematic uncertainties.

\bigskip
\noindent {\bf\boldmath{{Muon d$E$/d$x$}}:}
One of the main cuts in the event selection is the requirement that tracks
leave hits in at least 8 scintillator planes. Consequently, the event selection
efficiency is sensitive to the Monte Carlo simulation of muon energy loss. 
The simulation of muon energy loss depends on the underlying simulation 
of the physics processes
and the knowledge of the chemical composition of the MINOS detector. An 
uncertainty of 3\,\% in the muon range is assumed.  
These uncertainties result in systematic uncertainties 
of 1.7\,\% on the selection 
efficiency and 0.3\,\% on the up-down ratio.

\bigskip
\noindent {\bf{{Neutron Background}}:}
The background from cosmic-ray
induced neutrons has been studied using a GEANT 4 simulation\cite{bib:GEANT4} of muon
nuclear interactions in the rock and is found to be negligible.
In a Monte Carlo sample corresponding to four times the data exposure
no neutron event passed even the early stages of the event selection.

\section{Results}

From the $\livetime$ exposure considered in this paper,
$\Nobs$ candidate contained events are selected. 
The $\Nobs$ selected events are consistent with 
both the expectation of $\Nexpnoosc$ events assuming no neutrino
oscillations, and with the expectation of $\NexpSK$ events assuming 
$\Delta m^2_{23}=0.0024$\,eV$^2$ and $\Sinsq=1.0$. 
 The background contribution from 
cosmic-ray muons, $\Nback$, is obtained from data as described above. 
In addition, there is an expected background of $4.5\pm0.5$ from 
the combination of neutral current interactions and 
$\nue/\nuebar$ CC interactions.
The error in the MC expectation 
is dominated by the uncertainty on the neutrino
flux $\times$ interaction cross section which is estimated to be
$\NormSYS$. Table~\ref{tab:selectionsum} gives a breakdown of the
various contributions to the expected event rates.

\begin{table*}[htb]
\centering
\begin{tabular}{|l|c|c|c|c|c|c|c|}
\hline
Selection & Data & \multicolumn{6}{|c|}{Expectation (no oscillations)} \\ 
          &      & cosmic $\mu$ & $\nu_{\mu}/\bar{\nu}_{\mu}$ CC &  $\nue/\nuebar$ CC & NC  &   Rock $\nu_{\mu}$ &$\nu_{\tau}/\bar{\nu}_{\tau}$ CC\\ \hline 
 FC & $69$ & $   3.9\pm   0.4$ & $  81.2\pm   8.5$ & $   2.5\pm   0.3$ & $   2.0\pm   0.2$ & $   0.3\pm   0.1$ & $-$   \\ 
 PC Down & $25$ & $   0.6\pm   0.2$ & $  18.5\pm   1.9$ & $   0.1$ &   $-$ & $   0.1$ & $-$  \\ 
 PC Up & $13$ & $<0.36$ & $  17.4\pm   1.8$ & $-$ & $-$ & $   0.1$ & $-$ \\ \hline 
 Total & $107$ & $\Nback$ & $ 117.1\pm  12.2$ & $   2.6\pm   0.3$ & $   2.0\pm   0.2$ & $   0.5\pm   0.1$ & $-$  \\ \hline 
    &      & \multicolumn{6}{|c|}{Expectation ($\Delta{}m^2_{23} =0.0024\,\mbox{eV}^2$)} \\ 
 FC & $69$ & $   3.9\pm   0.4$ & $  58.4\pm   6.1$ & $   2.5\pm   0.3$ & $   2.0\pm   0.2$ & $   0.2$ & $   0.7\pm   0.1$  \\ 
 PC Down & $25$ & $   0.6\pm   0.2$ & $  17.5\pm   1.8$ & $   0.1$ & $-$ & $   0.1$ & $-$ \\
 PC Up & $13$ & $<0.36$ & $   9.2\pm   1.0$ & $-$ & $-$ & $   0.1$ &  $   0.5\pm   0.1$  \\ \hline  
 Total & $107$ & $\Nback$ & $  85.1\pm   8.9$ & $   2.6\pm   0.3$ & $   2.0\pm   0.2$ & $   0.4\pm   0.1$ & $   1.2\pm   0.1$  \\ \hline 
\end{tabular}
\caption{The numbers of data events in each selection category compared to the
expectation from different sources. The MC expectations from neutrino interactions
are given for both no oscillations and $\Sinsq=1.0$ and $\Deltamsq=0.0024\,\eV^2$.
The column referring to ``rock $\nu_{\mu}$'' refers to muons which are produced
by neutrino interactions in the surrounding rock. The cosmic muon backgrounds
in the FC and PC Down samples are estimated from data events passing all 
selection cuts with the exception of the veto shield. Entries marked as $-$ indicate
expectations of less than 0.05 of an event.  For the entries where no error is
quoted the error is less than 0.05 of an event.
\label{tab:selectionsum}}
\end{table*}

\begin{figure}
\includegraphics[width=\columnwidth]{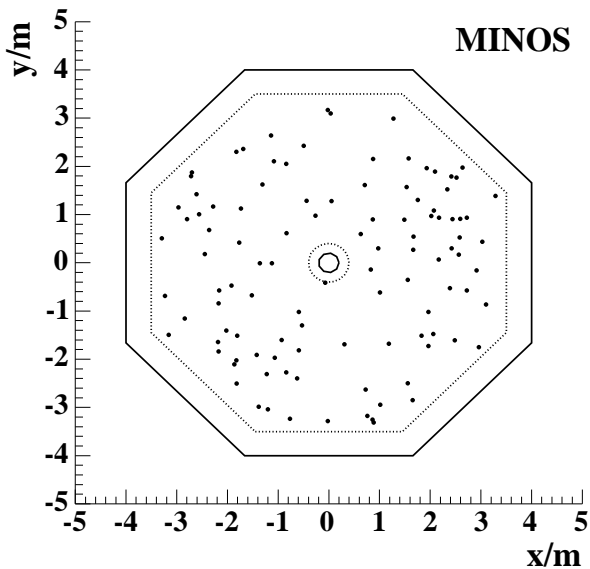}
 \caption{The reconstructed $x-y$ positions of the neutrino interaction vertices
          for the $\Nobs$ selected events. The vertex is defined as
          the start of the track, which is determined from timing. 
          The solid lines indicate the
          active region of the MINOS detector and the dotted lines
          indicate the boundaries of the fiducial volume.
 \label{fig:VertexPosition}}
\end{figure}

The $xy$ positions of the reconstructed neutrino
interaction vertices is shown in Figure~\ref{fig:VertexPosition}.
There is no evidence for a non-statistical accumulation of events 
in a particular region.

\begin{figure}
\includegraphics[width=\columnwidth]{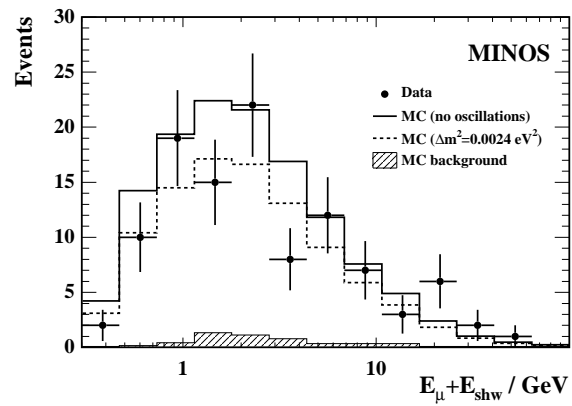}
 \caption{The reconstructed neutrino energy (logarithmic scale) 
          for the $\Nobs$ selected
          events compared to the MC expectation. The neutrino energy is
          taken to be the sum of the muon momentum and the
          energy of any hadronic shower associated with the assumed 
          interaction vertex. For the FC and PC samples, the muon momentum is 
          determined from range and curvature respectively.
          The solid histogram shows the MC expectation
          for the case of no neutrino oscillations, the  hatched histogram
          shows the cosmic-ray background and the points with error
          bars show the data. The dashed histogram shows the expectation 
          for $\numu\leftrightarrow\nutau$ oscillations with 
          $\Sinsq=1.0$ and $\Deltamsq=0.0024\,\eV^2$.
 \label{fig:Energy}}
\end{figure}

The reconstructed neutrino energy 
distribution of the $\Nobs$ candidate events is shown in Figure~\ref{fig:Energy}.
The neutrino energy is  calculated by summing the reconstructed
muon energy and the hadronic energy 
of any reconstructed shower associated with the start of the muon track.
For FC events, the muon energy is determined from the 
track range. For PC events the less precise momentum
from curvature is used.

The neutrino energy spectrum is sharply peaked towards lower energies and
the selected event sample is expected to have 
mean neutrino energy of 3.5\,GeV (2.0\,GeV for FC events and 
7.0\,GeV for the PC events) and mean muon energy of $2.4$\,\GeV. 
For low energy events the ability to determine the sense
of the muon track (up/down) is degraded. In MC, 96\,\%  of the selected 
events have the correct direction reconstruction. 
The remaining 4\,\% of
events not only have the wrong reconstructed sense, but as a 
consequence, are also assigned the incorrect charge. By requiring
that $|\RMSUP-\RMSDOWN|>0.66$\,ns (see Section~\ref{sect:reco})
and that the track traverses at least ten planes 
the fraction of mis-reconstructed events is reduced to 0.1\,\%.  
For the results that follow, the event sample is divided into two: 
a `Low Resolution' sample with $|\RMSUP-\RMSDOWN|<0.66$\,ns and events 
with `Good Timing' for which $|\RMSUP-\RMSDOWN|>0.66$\,ns. 
The numbers of events in
each category are listed in Table~\ref{tab:timing_cut}.
The $\NobsLowRes$ events classified as low resolution are
mainly short events and according to MC have a mean neutrino energy of 1.0\,\GeV.
For 85\,\% of the low resolution events, the muon is reconstructed
with the correct sense (up/down). However, in the oscillation analysis that follows,
the direction information from the low resolution sample is not used, due to
the significant fraction of events reconstructed with the wrong direction sense and the fact
that for this predominantly low energy sample, the mean angle between the incident
neutrino and final state muon is large.

\begin{table}[htb]
\centering
\begin{tabular}{|l|r|c|c|}
\hline
Selection       & Data & Expected        & Expected  \\ 
                &      & no oscillations & $\Deltamsq=0.0024\,\eV^2$ \\ \hline
Good Timing     & $\NobsGood$   & $\NexpnooscGood$    & $\NexpSKGood$  \\ 
Low Res.      & $\NobsLowRes$   & $\NexpnooscLowRes$    & $\NexpSKLowRes$  \\ \hline 
All Events      & $\Nobs$  & $\Nexpnoosc$    & $\NexpSK$  \\ 
\hline
\end{tabular}
\caption{Classification of events into samples with almost unambiguous direction
from timing (`Good Timing') and those where the direction from timing
is uncertain (`Low Resolution'). The errors are dominated by the systematic
 uncertainty in the neutrino event rate.  The MC expectations are given for both no oscillations and 
 $\Sinsq=1.0$ and $\Deltamsq=0.0024\,\eV^2$.}
\label{tab:timing_cut}
\end{table}

\begin{figure}
\includegraphics[width=\columnwidth]{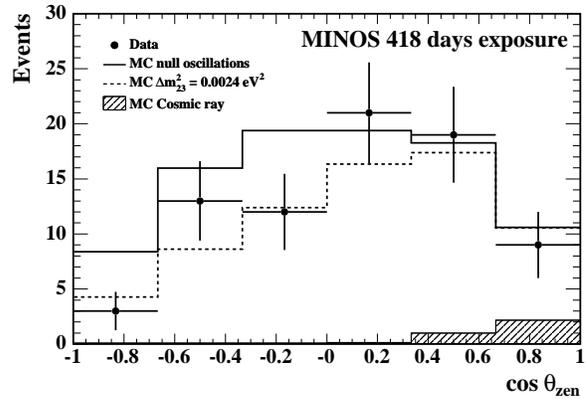}
 \caption{The reconstructed $\cos\thetazen$ distribution for the $\NobsGood$
          selected events with good timing compared to the MC expectation.
          The solid histogram shows the MC expectation
          for the case of no neutrino oscillations, the hatched histogram
          shows the cosmic-ray background and the points with error
          bars show the data. The dashed histogram shows the expectation 
          for $\numu\leftrightarrow\nutau$ oscillations with $\Sinsq=1.0$ and
          $\Deltamsq=0.0024\,\eV^2$. \label{fig:Zenith}}
\end{figure}

Figure~\ref{fig:Zenith} shows the reconstructed 
zenith angle distribution of the $\NobsGood$ 
candidate events with good timing. 
Of these events, $\NobsGoodDown$ 
are downward-going $(\coszen>0)$ and $\NobsGoodUp$ are
upward-going $(\coszen<0)$, giving a measured up-down ratio of
$0.57^{+0.17}_{-0.13}(stat.)$. The statistical errors correspond to the
68\,\% confidence interval calulated using Poisson statistics\cite{bib:Gehrels}.
The expected value from Monte Carlo
in the absence of neutrino oscillations is 
$0.92\pm0.03(sys.)$. The expected value is lower than one because of the 
different selection
efficiencies for upward- and downward-going events and the presence of background.
The upward-going/downward-going double ratio is:
\begin{eqnarray*}
 R_{\mathrm{up/down}}^{\mathrm{data}}/R_{\mathrm{up/down}}^{\mathrm{MC}} \!\!& = &\!\! 
   \rupdown^{+\rupdownstatplus}_{-\rupdownstatminus} (stat.) \pm \rupdownsys (sys.).
\end{eqnarray*}
This is approximately two standard deviations from unity, which is the 
expectation in the absence of neutrino oscillations. The systematic error 
is dominated by the experimental uncertainties; the estimated systematic
uncertainty on the predicted up/down neutrino flux ratio is less 
than 1\,\%\cite{bib:BarrPC} due to the relatively high energy of the 
selected neutrino events.

\subsection{Oscillation Analysis}

\begin{figure}
\includegraphics[width=\columnwidth]{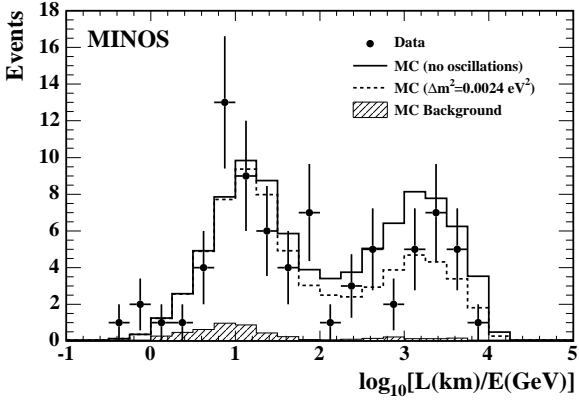}
 \caption{The reconstructed $\log_{10}{[ L(\mathrm{km})/E(\mathrm{GeV})]}$
          distribution compared to 
          the expectation. The solid histogram shows the MC expectation
          for the case of no neutrino oscillations, the hatched histogram
          shows the cosmic-ray background and the points with error
          bars show the data.
 The dashed histogram shows the expectation 
          for $\numu\leftrightarrow\nutau$ oscillations with $\Sinsq=1.0$ 
          and $\Deltamsq=0.0024\,\eV^2$.
 \label{fig:LOE}}
\end{figure}

In the two-flavour approximation, which is adequate for the level of
statistical precision considered here, the $\numu$ survival 
probability, $P(\numu\rightarrow\numu)$ is given by 
\begin{eqnarray*}
   P & = & 1.0-{\sin^2}\,{2\theta_{23}}\sin^2\left(1.27\Deltamsq[\mathrm{eV}^2].
    \frac{L[\mathrm{km}]}{E[\mathrm{GeV}]}\right),
\end{eqnarray*}
where $L$ is the distance traveled by the neutrino and $E$ is the neutrino energy.
The neutrino path length, $L$, is calculated from the reconstructed zenith 
angle assuming the neutrinos 
are produced at a height of $20$\,km in the Earth's atmosphere. 
Figure~\ref{fig:LOE} shows the reconstructed $L/E$ distribution for the
$\NobsGood$ candidate events with good timing. 

\begin{figure}
\includegraphics[width=\columnwidth]{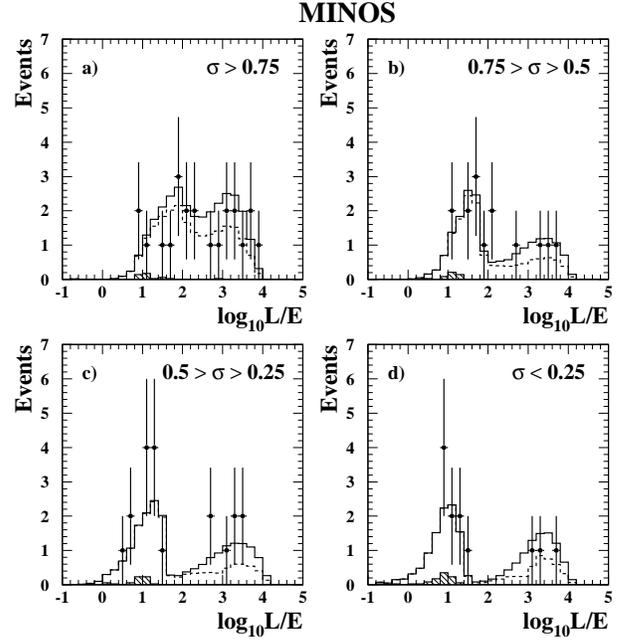}
 \caption{The reconstructed $\log_{10}{[ L(\mathrm{km})/E(\mathrm{GeV})]}$
          distribution binned in four regions of $\log{L/E}$ resolution, 
          $\sigma$.
          PC events with $|Q/p|/\sigma_{(Q/p)}<1$ are not used.
          The solid histogram shows the MC expectation
          for the case of no neutrino oscillations, the hatched histogram
          shows the cosmic-ray background and the points with error
          bars show the data. The dashed histogram shows the expectation 
          for $\numu\leftrightarrow\nutau$ oscillations with $\Sinsq=\bestSST$ 
          and $\Deltamsq=\bestDM$, the oscillation parameters corresponding to
          the best fit to the MINOS data.
 \label{fig:loevsrms}}
\end{figure}

The reconstructed $L/E$ distribution is used as the basis for
a fit to the hypothesis of $\numu\rightarrow\nutau$ oscillations.
The resolution on $L/E$ differs greatly event-to-event for
three main reasons: for PC events the muon momentum from
curvature may be poorly determined;
for low energy and/or high $y$ events 
the opening angle between the observed muon and the true neutrino
direction is large; and in the case where the muon direction 
is close to the plane defined by the horizon, relatively small changes 
in angle produce large changes in $L/E$. To address the first
issue, PC events with little observable curvature, 
$|Q/p|/\sigma_{(Q/p)}<1$, are not used in the fit to the $L/E$ distribution. 
To account for the different $L/E$ resolutions, in the oscillation
fit the data are binned according to $L/E$ resolution. To estimate
the event resolution a Bayesian 
approach has been adopted which allows the event-by-event 
$\log{(L/E)}$ probability density function (pdf) 
to be determined\cite{bib:Bayesfit}. The rms of the pdf, 
$\sigmaLoE$, gives a measure of the $\log{(L/E)}$ resolution of 
the event. For example, Figure~\ref{fig:loevsrms} shows the data 
binned in four regions of $\sigmaLoE$. For MC, the sensitivity of the 
$L/E$ distribution to neutrino oscillations increases with 
decreasing $\sigmaLoE$.

\subsubsection{Oscillation Analysis: Fit Procedure}

The selected events are divided into 10 equal sized bins 
of the estimated uncertainty in reconstructed $L/E$ ratio,
$\sigmaLoE$, ranging from $0.1-1.1$. Events
with $\sigmaLoE>1.0$ are included in the lowest resolution sample
and events with $\sigmaLoE<0.1$ are included in the highest resolution.  
A simultaneous fit is performed to the overall normalization
(using all selected events), the up-down ratio for $\NobsGood$
events with good timing, and separately the {\em shapes} of the upward and 
downward $L/E$ distributions for events with good timing 
and $|Q/p|/\sigma_{(Q/p)}>1$. In this way, each event is used only 
when the physical observable being fitted is well-measured.
A maximum likelihood 
fit to the data is 
performed using the negative log-likelihood function:
\begin{eqnarray*}
  -\ln{{\cal{L}}} = (\mu\!\!&-&\!\! N\ln{\mu})
 -\sum_k (N_u^k\ln{P_u^k} + N_d^k\ln{P_d^k}) \\
&-&\!\!\sum_{i_u} \ln{f_u^k([L/E]_{i_u})}
-\sum_{i_d} \ln{f_d^k([L/E]_{i_d})} \\ &+&\!\!\sum_j\frac{\alpha_j^2}{2\sigma^2_{\alpha_j}},
\end{eqnarray*} 
where $N$ is the total number of observed events and $\mu$ is
the total Monte Carlo expectation. The first two terms represent the Poisson probability
of observing $N$ events given the expectation of $\mu$. The normalization 
systematics
uncertainties are included as nuisance parameters (see below).
In the remaining terms, the superscript $k$
refers to the $k^{th}$ bin in $\sigmaLoE$. The sum
\begin{eqnarray*}
  \sum_k(N_u^k\ln{P_u^k}+N_d^k\ln{P_d^k})
\end{eqnarray*}
is the ``up-down'' likelihood. Here $N_u^k$ and
$N_d^k$ are the observed numbers of upward and 
downward-going events with good timing in bin $k$ 
of resolution; $P_u^k$ and $P_d^k$ are the Monte Carlo
probabilities that an event in resolution bin $k$ is upward or
downward going ($P_u^k+P_d^k=1$). The terms
\begin{eqnarray*} 
  \sum_{i_u} \ln{f_u^k( [L/E]_{i_u})} \ \ \ & \mathrm{and} & \ \ \ \
  \sum_{i_d} \ln{f_d^k( [L/E]_{i_d} )}
\end{eqnarray*}
are the likelihood functions for the observed $L/E$ distributions
of upward- and downward-going events respectively. Here the summations
are over the reconstructed upward and downward events respectively;  
$f_u^k( [L/E]_{i_u})$ is the normalized Monte Carlo pdf for 
the reconstructed $L/E$ distribution in the $k^{th}$ bin of
resolution (that of the event), 
evaluated at the measured value of $L/E$ of the event. 
The MC expectations for $\mu$, $P_u^k$, $P_d^k$, $f_u^k(L/E)$
and $f_d^k(L/E)$ 
include contributions from both neutrino interactions and cosmic-ray 
background and depend on ($\Deltamsq$, $\Sinsq$) and the
nuisance parameters representing the systematic
uncertainties. In calculating the expectations as a function of
oscillation parameters, the oscillation probabilities are
averaged over the distribution of neutrino production heights
obtained from the Bartol 3D model. 
In the fit, systematic effects are included using the nuisance
parameters, $\alpha_j$, which represent the deviation of a particular
parameter from its nominal value. The nuisance parameters contribute to the
likelihood function through the terms
\begin{eqnarray*} 
  \sum_j\frac{\alpha_j^2}{2\sigma^2_{\alpha_j}},
\end{eqnarray*} 
where $\sigma^2_{\alpha_j}$ is the estimated systematic uncertainty.
The following systematic effects are included: 
{\em i)} the uncertainty on the expected 
         neutrino event rate is taken to be $\pm10\,\%$;             
{\em ii)} a 3\,\% uncertainty on the muon momentum and a 5\,\% uncertainty on the
          hadronic energy scale; 
{\em iii)} a 3\,\% uncertainty on the relative efficiency for selecting upward-
          versus downward-going events; 
{\em iv)} to accommodate the uncertainty in the shape of the neutrino energy 
          spectrum, the spectrum is allowed to scale according to $1.0+0.1\beta(E_\nu-2)$ for 
          $E_\nu<2$\,GeV and $1.0+0.025\beta(E_\nu-2)$ for $E_\nu>2$\,GeV, where
          $\beta$ is normal distributed (these variations cover the differences
          in the neutrino energy spectra obtained from different flux 
          models\cite{bib:Barr,bib:Battistoni,bib:Bartol}); and 
{\em v)}  to allow for uncertainties in the modelling of neutrino cross sections,
          the relative cross section for quasi-elastic interactions is assumed to
          have a 20\,\% uncertainty. Since the neutrino flux times cross section 
          is normalized to Soudan~2 data, in the fit the systematic error associated 
          with the QE fraction only affects the shapes of the reconstructed $L/E$ 
          distributions.

With the exception of normalization, the systematic uncertainties have little 
impact on the resulting confidence regions. 
The above form of the likelihood function
is chosen to simplify the inclusion of systematic errors in the fit:
as normalization and up-down ratio are treated independently of shape.

\subsubsection{Oscillation Analysis: Results}

\begin{figure}
 \includegraphics[width=\columnwidth]{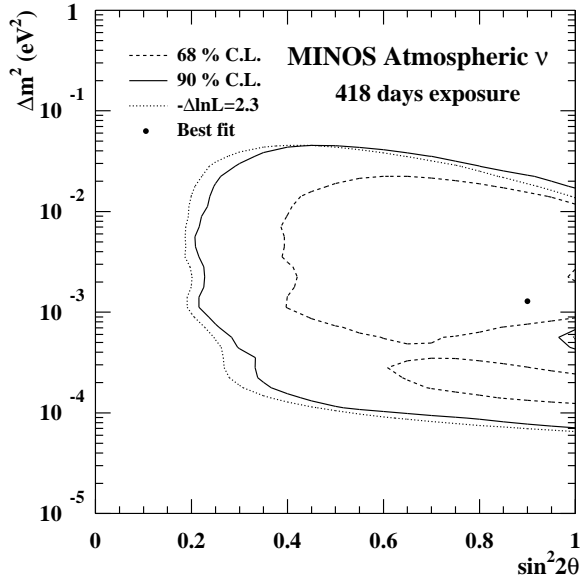}
 \caption{The 68\,\% and 90\,\% confidence limits on the oscillation parameters 
           obtained using the Feldman and Cousins approach. Also shown
           is the 90\,\% C.L. limit 
           calculated using $-\Delta{\ln{\cal{L}}}<2.3$.
 \label{fig:2DLnL}}
\end{figure}

For each hypothesized value of (\Deltamsq,\Sinsq) the negative
log-likelihood function
described above is minimized with respect to the nuisance parameters. 
For the data, the minimum likelihood occurs at $(\Deltamsq=\bestDM,\Sinsq=\bestSST)$. 
The 90\,\% confidence limits are obtained from the difference in the
log-likelihood function
$-\Delta{\ln{\cal{L}}} = {-\ln{\cal{L}}}(\Deltamsq,\Sinsq)+{\ln{\cal{L}}}_0$. 
Here $-\ln{{\cal{L}}_0}$ is the value of the negative log-likelihood function for the best fit to
the data. In the limit of Gaussian errors, the 90\,\% confidence level 
allowed regions of parameter space are defined by $-\Delta\ln{\cal{L}}<2.3$. 
The frequentist approach of Feldman and Cousins\cite{bib:FC} is used to determine
the value of $-\Delta{\ln{\cal{L}}}$ which corresponds to a particular confidence
level. For each point in parameter space, (\Deltamsqgen,\Sinsqgen), 
1000 Monte Carlo experiments are generated. In each experiment a value for each 
systematic bias is drawn from
a Gaussian distribution with standard deviation equal to the estimated 
systematic uncertainty. The generated data sample is fitted in the same manner as
the data and the value of $-\Delta{\ln{\cal{L}}}(\Deltamsqgen,\Sinsqgen)$ is
determined. 
The generated point is included in the 90\,\% confidence region 
if less than 90\,\% of the experiments yield a smaller value than obtained in the
data. For the MINOS data the 68\,\% and 90\,\% confidence limits obtained
using the Feldman and Cousins approach are shown in Figure~\ref{fig:2DLnL}. 
The 90\,\% C.L. contour is close to that obtained using 
$-\Delta\ln{\cal{L}}=2.3$. With the current statistics, the 
MINOS atmospheric neutrino data are consistent with a wide range of oscillation 
parameters including the most recent results from 
Super-Kamiokande\cite{bib:SuperK_new} and K2K\cite{bib:K2K}. The 
data disfavor the null oscillation hypothesis at the 
$\NullCons$ confidence level. 

\begin{figure}
\includegraphics[width=\columnwidth]{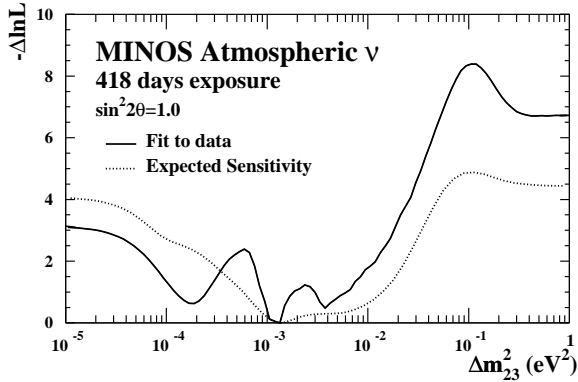}
 \caption{The $\Delta\cal{L}$ curve as a function of $\Deltamsq$ for the case
          of maximal mixing. The dotted curved shows the expected sensitivity 
         which is the average likelihood
         curve obtained from 10000 Monte Carlo experiments.
 \label{fig:1DLnL}}
\end{figure}

For completeness, Figure~\ref{fig:1DLnL} shows the 
likelihood as a function of $\Deltamsq$ for $\Sinsq=1.0$. The rises at large
and small values of $\Deltamsq$ are mainly due to the normalization and up-down
ratio. The structure within this broad minimum arises from the fit to the
shape of the $L/E$ distribution. The quality of the fit is good. As a 
measure of the fit quality, 
10000 simulated experiments were generated with 
$(\Deltamsq=0.0024\,\eV^2, \Sinsq=1.0)$ and the minimum value of 
$-{\ln{\cal{L}}}_0$ determined; in 84\,\% of these experiments the minimum
value of $-{\ln{\cal{L}}}_0$ exceeded that obtained from the fit to the data.
Figure~\ref{fig:1DLnL} also shows the expected sensitivity.

\subsection{Charge Ratio}

\begin{figure}
\includegraphics[width=\columnwidth]{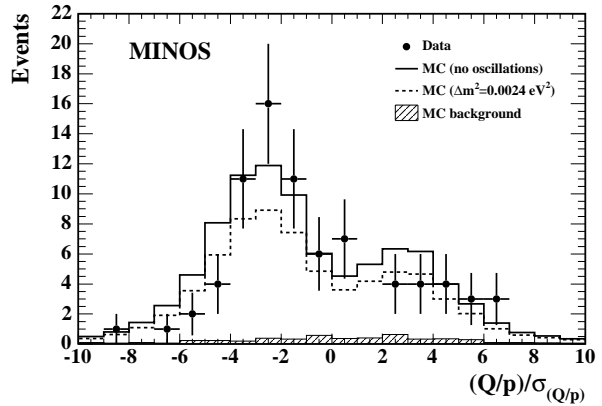}
 \caption{The reconstructed distribution of $(Q/p) / \sigma_{(Q/p)}$, the ratio of the 
          charge divided by momentum obtained from the track curvature divided by its error.
          The solid histogram indicates the Monte Carlo expectation
          assuming no oscillations, the hatched histogram
          shows the cosmic-ray background and the points with error
          bars show the data.
           The dashed histogram shows the expectation 
          for $\numu\leftrightarrow\nutau$ oscillations with $\Sinsq=1.0$ 
          and $\Deltamsq=0.0024\,\eV^2$.
  \label{fig:sigmaqp}}
\end{figure}

The selected contained events with unambiguous timing information are
divided into neutrino and anti-neutrino interactions
on the basis of the reconstructed muon charge obtained from 
the curvature of the reconstructed muon track. Only events with 
unambiguous direction from timing are considered;
events with an incorrect direction will be reconstructed with
the wrong zenith angle and will have their charge inverted. 
Figure~\ref{fig:sigmaqp}
shows the distribution of $(Q/p)/\sigma_{(Q/p)}$ for the $\NobsGood$ events
with well determined direction from timing compared to the MC expectation.
Events are classified as: $\numu$ for 
$(Q/p) / \sigma_{(Q/p)} < -2$; 
$\numubar$ for
$(Q/p) / \sigma_{(Q/p)} > +2$; 
or events which are ambiguous,
$|(Q/p) / \sigma_{(Q/p)}| \le 2$. 
The selected numbers of events in each charge category are compared to the MC
expectations in Table~\ref{tab:data_q}.  

\begin{table}[htb]
\centering
\begin{tabular}{|l|c|c|c|}
\hline
Selection       & Data & Expected        & Expected  \\ 
                &      & no oscillations & $\Deltamsq=0.0024\,\eV^2$ \\ \hline
Low Res.  & $\NobsLowRes$   & $\NexpnooscLowRes$    & $\NexpSKLowRes$  \\ 
Ambig. $\numu$/$\numubar$ & $\NobsQambig$ & $\NexpnooscQambig$ & $\NexpSKQambig$ \\
$\numu$        & $\Nobsnu$      & $\Nexpnunoosc$       & $\NexpnuSK$  \\ 
$\numubar$      & $\Nobsnubar$  & $\Nexpnubarnoosc$    & $\NexpnubarSK$  \\ 
\hline
\end{tabular}
\caption{Event classification according to timing and track curvature.
The four categories are 
events with ambiguous direction from timing (`Low Resolution'), 
events with good timing information but ambiguous charge assignment
(`Ambiguous $\numu$/$\numubar$'), $\numu$ and $\numubar$. 
 The errors are dominated by the  systematic
 uncertainty in the neutrino flux $\times$ cross section. 
 The MC expectations are given for both no oscillations and 
 $\Sinsq=1.0$ and $\Deltamsq=0.0024\,\eV^2$.}
\label{tab:data_q}
\end{table}

Of the events where it is possible to
cleanly tag the charge of the muon,
18 are identified as $\numubar$ candidates and 34 as $\numu$ candidates,
yielding a $\numubar$ to $\numu$ ratio of:
\begin{eqnarray*}
R^{\mathrm{ data}}_{\numubar/\numu} \!\!&=&\!\! \nunubarrat^{+\nunubarratstatplus}_{-\nunubarratstatminus}\mathrm(stat.)\pm\nunubarratsys\mathrm(sys.). 
\end{eqnarray*}
The systematic uncertainty is the experimental uncertainty associated with charge
identification.  The uncertainty was estimated by
shifting and smearing the Monte Carlo reconstructed values of 
$(Q/p) / \sigma_{(Q/p)}$ whilst maintaining reasonable agreement 
between data and Monte Carlo for the stopping muon data shown in 
Figure~\ref{fig:StoppersCharge}.
For the purposes of studying possible biases in the charge reconstruction, 
37\,\% of the data were recorded with the coil current
reversed.  Consistent values for the $\numubar$/$\numu$ ratio are found in
the normal and reversed current data samples; 22(12) events are identified as $\numu$ and
12(6) are identified as $\numubar$ in the normal(reversed) field data samples.
From Monte Carlo, the expected ratio of identified $\numubar$ to $\numu$ events 
is $0.550$, where it is assumed that both neutrinos and anti-neutrinos oscillate 
with the same parameters. This expected ratio
is almost independent of the values of the oscillation parameters provided they
are the same for neutrinos and anti-neutrinos. 
The ratio of $\numubar$ to $\numu$ events in the data compared to the
Monte Carlo expectation (Bartol 3D and NEUGEN3) 
assuming the same oscillation parameters for
neutrinos and anti-neutrinos is: 
\begin{eqnarray*}
 R^{\mathrm{data}}_{\numubar/\numu}/R^{\mathrm{MC}}_{\numubar/\numu}
 & = & 
   \rnubarnu^{+\rnubarnustatplus}_{-\rnubarnustatminus} (stat.) \pm \rnubarnusys (sys.).
\end{eqnarray*}
The statistical errors correspond to the
68\,\% confidence interval calulated using Poisson statistics\cite{bib:Gehrels}.
The systematic error includes the experimental uncertainty associated with the
muon charge identification (\rnubarnuexpsys), the uncertainty in the
relative $\numubar$ to $\numu$ flux 
(\rnubarnufluxsys) and the relative uncertainty in the $\numu$ to $\numubar$ cross section
(\rnubarnuxsecsys).
The systematic errors on the relative $\numubar$ to $\numu$ fluxes and cross sections were
estimated taking into account the energy spectra of the charge-tagged neutrino and anti-neutrino
events. From Monte Carlo, the sample of events where the charge of muon is cleanly identified 
has a mean neutrino energy of 3.7\,GeV, with 95\,\% of the expected events having 
neutrino energies between 1\,GeV and 10\,GeV. In Monte Carlo, 40\,\% of these
events  arise from  quasi-elastic interactions, 30\,\% arise from resonance production, 
and 30\,\% from deep inelastic scattering.  
In this energy range the uncertainty 
on the ratio of atmospheric $\numubar$ to $\numu$ flux was estimated to be 
8\,\%\cite{bib:Gaisser} for the Bartol 1D model. Recent studies based on the Bartol 3D model  
give an estimated uncertainty of 4\,\%\cite{bib:BarrPC}. 
Since there is limited data on anti-neutrino cross sections in the
energy range $1-5$\,\GeV\cite{bib:Erriquez, bib:Fanourakis, bib:Anikeev}, 
the variation in the predicted $\numubar$/$\numu$ event rate
was studied through conservative changes to the neutrino cross section 
model. The variations considered include parameters affecting the
free nucleon cross sections such as axial vector masses, choice of
PDF set, and model  for the resonance region. Similarly the 
effect of changes to the nuclear
physics model which affect the rate via  Pauli blocking of quasi-elastic
and nuclear shadowing of DIS events was studied.  
The quadrature sum of these changes is 13.5\,\%.
The largest contribution to this uncertainty comes from the treatment
of resonance production and the  resonance/DIS transition region.
A 12\,\% difference is found in comparing a model which explicitly includes
resonance production \cite{bib:NEUGEN, bib:Rein-Seghal} versus one which
uses a QCD-based approach \cite{bib:Bodek-Yang}.   The size of this difference
ultimately reflects the uncertainty in the experimental data to which
these models are tuned.

\begin{figure*}
\includegraphics[width=\textwidth]{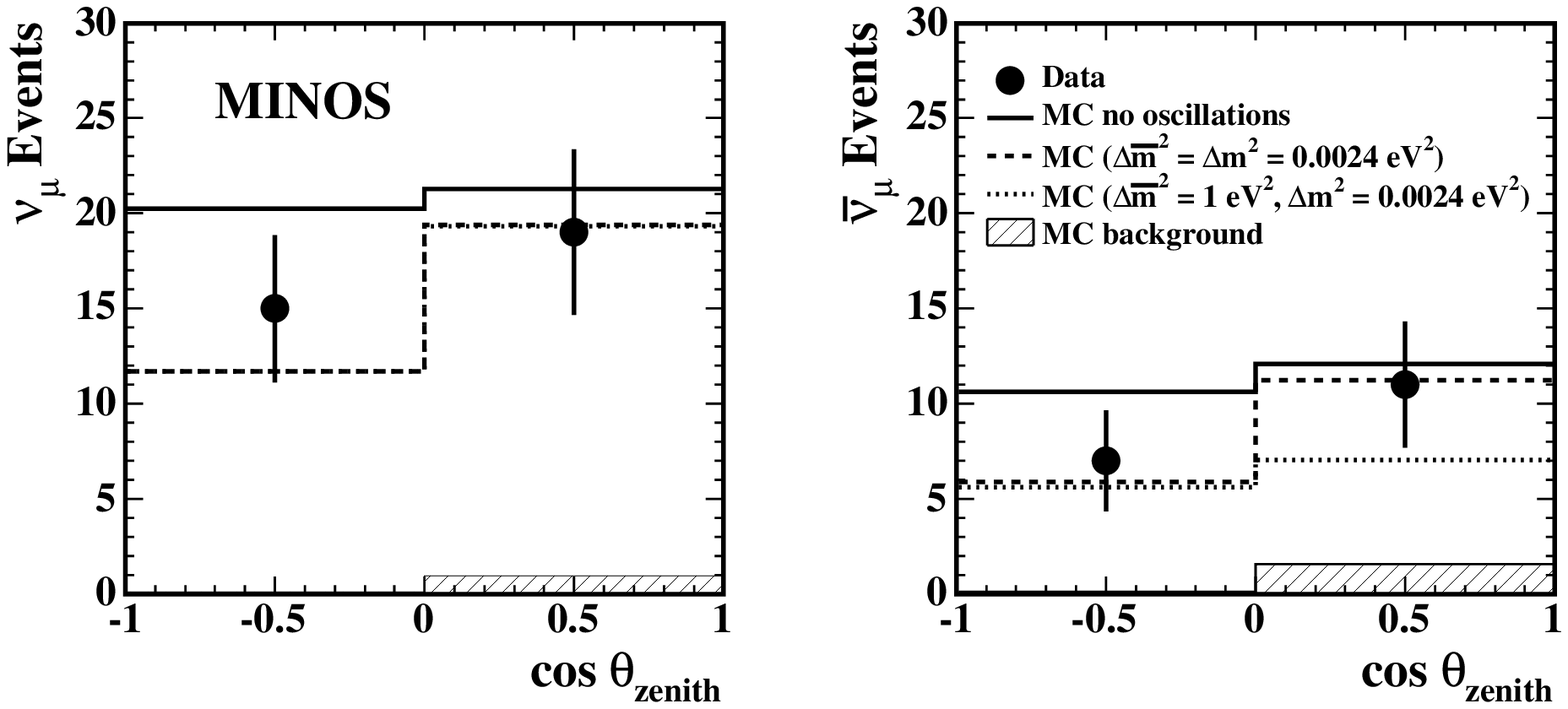}
 \caption{The charge-separated up/down distributions of the events which are cleanly identified 
          as $\numu$ or $\numubar$. The data ($\numu$ on the left and $\numubar$ on the right)
 are compared to the expectation for i) no oscillations; ii) the case
where both $\numu$ and $\numubar$ oscillate with $\Deltamsq=0.0024\,\eV^2$ 
and; iii) the case where $\numu$ oscillate with $\Deltamsq=0.0024\,\eV^2$ and 
$\numubar$ oscillate with $\Deltamsq=1.0\,\eV^2$. For the oscillated predictions
maximal mixing is assumed for $\numu$ and $\numubar$.
 \label{fig:Qzenith}}
\end{figure*}

CPT violating models which attempt to explain the LSND data suggest a large value of
$\Deltamsq$ for anti-neutrinos\cite{bib:CPT1,bib:CPT2}. 
In principle, the MINOS data will be able to address
this possibility by measuring the oscillation parameters for the selected anti-neutrino
sample. Figure~\ref{fig:Qzenith} shows up/down distribution of the 18 $\numubar$ and
34 $\numu$ events compared to the expectation for: i) no oscillations; ii) the case
where both $\numu$ and $\numubar$ oscillate with $\Deltamsq=0.0024\,\eV^2$ (maximal mixing) 
and; iii) the case where $\numu$ oscillate with $\Deltamsq=0.0024\,\eV^2$ (maximal mixing) and 
$\numubar$ oscillate with $\Deltamsq=1.0\,\eV^2$ (maximal mixing). The data are consistent
with the same oscillation parameters for neutrinos and anti-neutrinos. However, with the
current statistics the possibility of a large value of  $\Deltamsq$ for anti-neutrinos
cannot be excluded.

\section{Summary}

The MINOS far detector has been taking data since the beginning of August 
2003 at a depth of 2070 meters water-equivalent in the Soudan mine, 
Minnesota. This paper presents the first MINOS observations of \numu\ and 
\numubar\ charged current atmospheric neutrino interactions based on an 
exposure of $\livetime$. A total of $\Nobs$ candidate 
contained-vertex neutrino interactions is observed, consistent with both 
the expectation of $\Nexpnoosc$ for no neutrino oscillations and 
$\NexpSK$ for $\Deltamsq=0.0024$\,eV$^2$ and $\Sinsq=1.0$. The expected 
numbers of events include the estimated background from cosmic-ray muons,
$\Nback$, obtained from data. The errors on the expectation are dominated 
by $\NormSYS$ uncertainty on the neutrino event rate which
was obtained using results from the Soudan~2 collaboration.
Of the events for which the direction can be cleanly identified, the 
ratio of upward to downward-going events in the data is compared to the
Monte Carlo expectation in the absence of neutrino oscillations giving:
\begin{eqnarray*}
 R_{\mathrm{up/down}}^{\mathrm{data}}/R_{\mathrm{up/down}}^{\mathrm{MC}} 
 & = & 
   \rupdown^{+\rupdownstatplus}_{-\rupdownstatminus} (stat.) \pm \rupdownsys (sys.).
\end{eqnarray*}
An extended maximum likelihood fit to the observed $\log{L/E}$ distribution 
yields a best fit value of 
$(\Deltamsq=\bestDM,\Sinsq=\bestSST)$ and 90\,\% confidence limits of
$(\lowDM<\Deltamsq<\highDM,\Sinsq>\lowSST)$. 
The consistency of the data with the null hypothesis of no neutrino
 oscillations is investigated;
the data exclude the null hypothesis at the $\NullCons$ confidence level.

The curvature of the observed muons in the \Bfield\ MINOS magnetic field 
is used to separate $\numu$ and $\numubar$ interactions. 
Of the selected events for which it is possible to
cleanly determine the charge of the muon, $\Nobsnubar$ are identified 
as $\numubar$ candidates and $\Nobsnu$ as $\numu$ candidates, 
giving an observed $\numubar$ to $\numu$ ratio  of
$\nunubarrat^{+\nunubarratstatplus}_{-\nunubarratstatminus}\mathrm(stat.)\pm\nunubarratsys\mathrm(sys.)$. 
The fraction of $\numubar$ events in the data is compared to the
Monte Carlo expectation assuming neutrinos and anti-neutrinos oscillate in same
manner giving: 
\begin{eqnarray*}
 R^{\mathrm{data}}_{\numubar/\numu}/R^{\mathrm{MC}}_{\numubar/\numu}
 & = & 
   \rnubarnu^{+\rnubarnustatplus}_{-\rnubarnustatminus} (stat.) \pm \rnubarnusys (sys.).
\end{eqnarray*}
Although the statistics are limited, this is the first direct observation of  
atmospheric neutrino interactions separately for $\numu$ and $\numubar$.
The data are consistent with neutrinos and anti-neutrinos oscillating with the 
same parameters, although CPT violating scenarios with large values of
$\Deltamsq$ for anti-neutrinos are not excluded with the current data.

\begin{acknowledgments}

This work was supported by the U.S. Department of Energy, the U.K. Particle
Physics and Astronomy Research Council, the U.S. National Science Foundation, 
the State and University of Minnesota, the Office of Special Accounts for
Research Grants of the University of Athens, Greece, and FAPESP (Fundacao de
Amparo a Pesquisa do Estado de Sao Paulo) and CNPq (Conselho Nacional de
Desenvolvimento Cientifico e Tecnologico) in Brazil.  We
gratefully acknowledge the Minnesota Department of Natural Resources for their
assistance and for allowing us access to the facilities of the Soudan
Underground Mine State Park. We also thank the crew of the Soudan Underground
Physics Laboratory for their tireless work in building and operating the MINOS
far detector.

\end{acknowledgments}

\end{document}